\documentclass[12pt]{iopart}

\usepackage{graphicx} 
\usepackage{amssymb}
\usepackage{xcolor}
\begin{document}

\title{Correlations and transport in exclusion processes with general finite memory}

\author{Eial Teomy and Ralf Metzler}

\address{Institute for Physics \& Astronomy, University of Potsdam, Karl-Liebknecht-Stra{\ss}ße 24/25, D-14476 Potsdam-Golm, Germany}
\ead{eialteom@gmail.com}

\begin{abstract}
We consider the correlations and the hydrodynamic description of random walkers with a general finite memory moving on a $d$ dimensional hypercubic lattice. We derive a drift-diffusion equation and identify a memory-dependent critical density. Above the critical density, the effective diffusion coefficient decreases with the particles' propensity to move forward and below the critical density it increases with their propensity to move forward. If the correlations are neglected the critical density is exactly $1/2$. We also derive a low-density approximation for the same time correlations between different sites. We perform simulations on a one-dimensional system with one-step memory and find good agreement between our analytical derivation and the numerical results. We also consider the previously unexplored special case of totally anti-persistent particles. Generally, the correlation length converges to a finite value. However in the special case of totally anti-persistent particles and density $1/2$, the correlation length diverges with time. Furthermore, connecting a system of totally anti-persistent particles to external particle reservoirs creates a new phenomenon: In almost all systems, regardless of the precise details of the microscopic dynamics, when a system is connected to a reservoir, the mean density of particle at the edge is the same as the reservoir following the zeroth law of thermodynamics. In a totally anti-persistent system, however, the density at the edge is always higher than in the reservoir. We find a qualitative description of this phenomenon which agrees reasonably well with the numerics.

\end{abstract}

\noindent{\it Keywords\/}: Exclusion process, persistence, lattice gas, memory, random walk

\maketitle

\section{Introduction}

The active and passive motion of biological cells and the motion of their components inside them is a complicated out of equilibrium process which occurs due to many factors, some of them still unknown \cite{ref1}. This motion has been investigated at the single-body level \cite{Hasnain2015,ref2,ref3}, many-body level \cite{ref4,Berthier2013,ref5,Zimmermann2016,ref6,Reichhardt2014,Zachreson2017,Reichhardt2014b,Lam2015,Graf2017,Illien2015}, and continuum level \cite{ref7}. At the many-body level the focus is mostly on the interactions between cells or bacteria, be they hydrodynamic \cite{ref4}, mutually aligning as in the Viscek model \cite{ref5,Zimmermann2016}, energetic \cite{Zimmermann2016,ref6,Reichhardt2014,Zachreson2017}, or steric \cite{Reichhardt2014b,Lam2015,Graf2017,Illien2015,Fisher2014}. 

The motions of individual cells or bacteria are modelled in various ways, which can be thought of as a random walk with a certain type of memory. One of the most common models, motivated by experimental observations \cite{Berg1990}, is a run and tumble motion \cite{ref2,Reichhardt2014}, in which the walker moves in a straight line for some time, and then abruptly changes its direction. This model is captured by a memory term which leads to an increased probability of turning as more time passes since the last turn. A twitching motion \cite{Zachreson2017} or motion with a self aligning director \cite{Lam2015} is captured by a one-step memory term, i.e. the velocity at each step depends on the velocity in the previous step but not on longer-reaching memory terms. Other biological processes are also described as random walks with memory \cite{Schulz2011,Ghosh2015b,Hermann2017}.

In random walks with memory, each step the walker makes depends not only on its location in the previous step but on its history. It might depend on its entire history, or a finite part of it. For example, in one of the first random walk models that included memory \cite{Taylor1921}, a single walker moves on a one-dimensional lattice. At each step, the walker either moves in the same direction as it did in the previous step with probability $\frac{1}{2}+\delta$, or in the opposite direction with probability $\frac{1}{2}-\delta$. This rule mimics inertia, and does not introduce any global bias in any specific direction. The basic random walk model is retrieved for $\delta=0$. Such walkers with one-step memory are also called persistent walkers. Since the introduction of this model, it was expanded in various forms to explain different phenomena in various fields, such as polymer chains \cite{Tchen1952}, animal movement \cite{Kareiva1983}, scattering in disordered media \cite{Boguna1998}, motion of bacteria \cite{Hasnain2015}, artificial microswimmers \cite{Romanczuk2012,Ghosh2015}, and motion in ordered media \cite{Tahir}.

A different class of random walk models emulates the interactions in many-body systems. In these models, called lattice-gas models, many walkers move on a discrete graph or lattice with some type of interaction between the different particles. In the Simple Symmetric Exclusion Principle (SSEP) model \cite{Spitzer1970} the interaction is purely steric. Each site on a lattice is either vacant or occupied by at most one walker, and each walker has an internal clock, independent of the other walkers, which governs the timing of its attempted moves. If a walker attempts to move to an already occupied site, it remains in place. In the Asymmetric Simple Exclusion Principle (ASEP) model \cite{Spitzer1970}, the walkers are biased to move in a certain direction, and it has been used to describe transport phenomena in biology \cite{Graf2017,Melbinger2011}. A special consideration is given to one-dimensional systems \cite{ref8}, which emulates transport along a narrow channel, such as transport of water \cite{Waghe2012} or drugs \cite{Yang2010} through nanotubes, or of molecular motors in cellular protrusions \cite{Graf2017} and along microtubules \cite{Melbinger2011,ref11}. The single file diffusion in one-dimensional systems is known to be anomalous, even without memory \cite{Harris1965,Sanders2014}. The basic SSEP and ASEP models have been expanded to include energetic interactions \cite{Spohn1983}, a single biased particle surrounded by unbiased particles \cite{ref10}, birth and death of particles \cite{Markham2013}, higher site occupancy \cite{Arita2014}, spatial inhomogeneities \cite{Szavitz2018} and kinetic constraints \cite{Ritort2003}.

There are several studies that combine these two variations of the basic random walk, and they investigate three characteristics of this type of models. First, this model may be considered as a coarse-grained version of active Brownian particles (ABP) \cite{Romanczuk2012}, and it was shown that it indeed shows motility induced phase separation \cite{Whitelam2017,Soto2014}, one of the hallmarks of ABP. Second, some studies derived an effective hydrodynamic description in either one-dimensional \cite{Treloar2011,Kourbane2018} or higher-dimensional \cite{Manacorda2017,Gavagnin2018} systems, including anomalous walkers \cite{Arita2018}. The third group of studies investigates the mean squared displacement (MSD) of crowded walkers with memory, in particular the short time approximation of the MSD \cite{Galanti2013}, the MSD of interacting subdiffusive random walkers in a one-dimensional system \cite{Sanders2014}, the MSD in the very high density limit in one-dimension \cite{Bertrand2018}, and the effective diffusion coefficient of a cross-shaped persistent walker in a bath of memory-less cross-shaped walkers \cite{Chatterjee2018}.

In this paper we generalise our previous study \cite{Teomy2019} and consider the correlations and the hydrodynamic description of random walkers with a general finite memory moving on a $d$ dimensional hypercubic lattice. If the velocity autocorrelations are positive, we call the walkers persistent, while if they are negative we call them anti-persistent. We derive a drift-diffusion equation which takes the non-negligible correlations between the particles into account. We identify a memory-dependent critical density which governs the difference between the density-dependent bulk diffusion coefficient $D$ from the memory-less one $D_{0}$. For persistent walkers, below the critical density $D>D_{0}$ while above it $D<D_{0}$. For anti-persistent walkers, the situation is reversed: below the critical density $D<D_{0}$ while above it $D>D_{0}$. If the correlations are neglected the critical density is exactly $1/2$. We also derive a low-density approximation for the same time correlations between different sites, again for a general finite memory on a $d$ dimensional hypercubic lattice. We perform extensive simulations in a one-dimensional system with one-step memory and find excellent agreement between our analytical derivation and the numerical results. 

Finally, we also consider the previously unexplored special case of totally anti-persistent particles. Generally, the correlations converge to their steady state values after a finite time and have a finite correlation length. However in the special case of totally anti-persistent particles and density $1/2$, the correlations do not converge and the correlation length diverges with time. Furthermore, connecting a system of totally anti-persistent particles to external particle reservoirs creates a new phenomenon: In almost all systems, regardless of the precise details of the microscopic dynamics, when a system is connected to a reservoir, the mean density of particles at the edge is the same as in the reservoir following the zeroth law of thermodynamics. In a totally anti-persistent system, however, the density at the edge is always higher than in the reservoir.

The details of the model we investigate are described in section \ref{sec_model}. Section \ref{sec_diffusion} is devoted to the derivation of an effective diffusion equation for the coarsed-grained density. In section \ref{sec_corr} we look at the correlations between the states of two different particles. Section \ref{sec_numerical} contains a comparison between our analytical results and the numerical simulations. The special case of total anti-persistence is covered in section \ref{sec_tap}. Finally, section \ref{sec_summary} summarises the paper.

\section{Description of the model}
\label{sec_model}
We consider a lattice gas on a $d$-dimensional hypercubic lattice. Each site on the lattice can be either vacant or occupied by at most one particle. Each particle has an independent exponential clock with mean time $\tau$. When the clock rings, the particle attempts to move to one of its $2d$ nearest neighbours. If the target site is vacant, the particle moves. Otherwise, it remains in place. In both cases, its clock resets. 

\begin{figure}
\centering
\includegraphics[width=0.6\columnwidth]{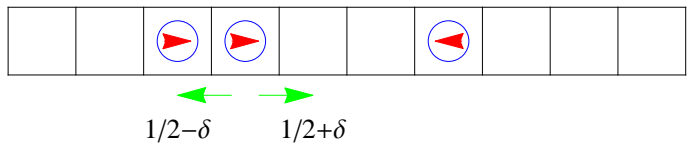}
\caption{An illustration of the one-step one-dimensional model. The last direction in which the particle moved is denoted by the red arrow inside the circle. At each step the particle turns in one of the directions with probabilities shown near the green arrows, and moves in that direction if the target site is vacant.}
\label{1dmodel}
\end{figure}

The target direction, however, is not chosen from a uniform distribution but it rather depends on the history of the particle. As a simple example, consider particles with one-step memory moving on a one-dimensional (1D) lattice, as illustrated in figure \ref{1dmodel}. The probability that a particle attempts to move in the same direction as in its previous state is $\frac{1}{2}+\delta$ with $-\frac{1}{2}\leq\delta\leq\frac{1}{2}$, and the probability it reverses its direction is $\frac{1}{2}-\delta$. We call the parameter $\delta$ the persistence parameter, since it encodes the tendency of the particle to persist in its motion. Note that since the probability distribution for choosing the direction of motion is relative to the current direction of motion, there is no global bias in the system unless it is imposed from the boundaries.

More generally, we may consider particles with $m$-step memory, i.e. that the probability distribution of the attempted direction of motion depends on the directions in which the particle attempted to move in its previous $m$ steps. Each particle may therefore be in one of $(2d)^{m}$ states which encodes its memory, and the transition probabilities between the states is given by the entries of the matrix ${\cal M}$. Even more generally, we may consider particles with infinite memory, and the transition matrix ${\cal M}$ is more accurately called an operator. Any type of previously investigated non-biased lattice-based model may be presented using this formulation, for example the elephant walk model \cite{Arita2018,Schutz2004,Coletti2017} or a run and tumble motion \cite{ref2,Reichhardt2014}. 

Although \textit{a priori} it may not appear so, this model is in fact Markovian in the following sense \cite{Othmer1988}. We may expand the phase space such that each state is defined by $L^{d}$ variables, representing the sites of the lattice, each having one of $1 + (2d)^{m}$ values: vacant or occupied with a specific memory. The $2d$ factor accounts for the $2d$ directions in which the particle could have moved in each of the previous $m$ steps. The transition rates between those states do not depend on the memory of the process in this expanded phase space. This also holds for models with infinite memory \cite{Berbee1987}.

Note that although the net current is zero, this model is out of equilibrium because it does not obey detailed balance. Consider for example a particle moving to the vacant site to its right, and that in its previous step it also moved to the right. Such a move occurs with some finite
probability depending on the exact form of the memory term. The opposite transition, however, has a zero probability of occurring, since if the particle moves to the now vacant adjacent site to its left its last move was to the left, and it is thus in a different state than the one it started from. We will call the situation at which there is no external force a ``pseudo-equilibrium''.

Our analytical results consider mostly models with a general finite memory term, but they are also relevant for models with infinite memory terms with correlations that decay fast enough. In the numerical results that follow, and also in some of the analytical derivations, for simplicity we consider one-step memory models in one dimension. 

\section{Effective diffusion equation}
\label{sec_diffusion}
We consider walkers with a finite general isotropic memory term. We denote by \textcolor{black}{$\eta=\left(\mathbf{\eta_{1}},\mathbf{\eta_{2}},...\right)$} the memory of the particle. \textcolor{black}{$\mathbf{\eta}_{n}$ denotes the direction to which the walker attempted to move in its $n$'th previous step}, such that $\mathbf{\eta}_{1}$ is the last step. The probability that a particle with memory $\eta'$ attempts to move such that its new memory is $\eta$ is given by the matrix element ${\cal M}_{\eta,\eta'}$. The probability that site $\mathbf{r}$ is occupied by a particle with history $\eta$, $P(\mathbf{r},\eta)$ is governed by the evolution equation
\begin{eqnarray}
\tau\frac{\partial P(\mathbf{r},\eta)}{\partial t}=-P(\mathbf{r},\eta)+\sum_{\eta'}P(\mathbf{r},\eta';\mathbf{r}+\mathbf{\eta}_{1}){\cal M}_{\eta,\eta'}+\nonumber\\
+\sum_{\eta'}{\cal M}_{\eta,\eta'}\left[P(\mathbf{r}-\mathbf{\eta}_{1},\eta')-P(\mathbf{r}-\mathbf{\eta}_{1},\eta';\mathbf{r})\right] ,\label{eq1}
\end{eqnarray}
\textcolor{black}{where $P\left(\mathbf{r},\eta;\mathbf{r}'\right)$ is the probability that site $\mathbf{r}$ is occupied by a particle with memory $\eta$ and site $\mathbf{r}'$ is occupied. 
The four terms on the right hand side of (\ref{eq1}) correspond to the following processes: the particle did not attempt to move; the particle in site $\mathbf{r}$ previously had memory $\eta'$, attempted to move in direction $\mathbf{\eta}_{1}$ but site $\mathbf{r}+\mathbf{\eta}_{1}$ is occupied; the particle in site $\mathbf{r}-\mathbf{\eta}_{1}$ previously had memory $\eta'$, and it attempted to move to site $\mathbf{r}$; and the particle in site $\mathbf{r}-\mathbf{\eta}_{1}$ previously had memory $\eta'$, and it attempted to move to site $\mathbf{r}$ but failed because site $\mathbf{r}$ is occupied.}

Taking the hydrodynamic limit, we find that the total occupancy probability $P\left(\mathbf{r}\right)$, defined by
\begin{eqnarray}
P\left(\mathbf{r}\right)=\sum_{\eta}P\left(\mathbf{r},\eta\right) ,
\end{eqnarray}
satisfies the drift-diffusion equation
\begin{eqnarray}
\frac{\partial P}{\partial t}=\sum_{\mathbf{d},\mathbf{d}'}\frac{\partial}{\partial\mathbf{d}}\left[D_{\mathbf{d},\mathbf{d}'}\left(P\right)\frac{\partial P}{\partial\mathbf{d}'}+v_{\mathbf{d}}(P)P\right] ,
\end{eqnarray}
with
\begin{eqnarray}
&D_{\mathbf{d},\mathbf{d}'}(P)=D_{0}\left[\delta_{\mathbf{d},\mathbf{d}'}+4c\left(1-P\right)\left(1-2P-\frac{\partial C_{1}}{\partial P}\right)\right] ,\nonumber\\
&\textcolor{black}{v_{\mathbf{d}}(P)=\frac{2a}{\tau}\frac{C_{2}}{P}} ,
\end{eqnarray}
\textcolor{black}{where $\mathbf{d}$ denotes the $d$ directions ($\hat{x},\hat{y},...$),} $D_{0}$ is the diffusion coefficient in a memory-less system, $c$ is a constant which depends on the properties of the matrix ${\cal M}$, and $C_{1}$ and $C_{2}$ are correlations between the histories of particles in adjacent sites. See Appendix \ref{app_diff} for more details and the full derivation. The constant $c$ is positive if the velocity autocorrelations are positive (i.e. the particles are persistent) and is negative if the velocity autocorrelations are negative (i.e. the particles are antipersistent). \textcolor{black}{This correction to the base diffusion coefficient is qualitatively similar to the combined effect of persistence and finite density on the MSD, which increases with density for highly anti-persistent walkers \cite{Teomy2019}}. Note that as this is not a gradient model, using the density dependence of the correlation functions at pseudo-equilibrium is only an approximation \cite{Teomy2017}. From symmetry, we find that at pseudo-equilibrium $C_{2}=0$. 

In the simplest case of particles with one-step memory moving on a one dimensional lattice we find that
\begin{eqnarray}
&c=\frac{\delta}{1-2\delta} ,\nonumber\\
&C_{1}=C_{0}+2\delta\left(C_{+-}-C_{-+}\right) ,\nonumber\\
&C_{2}=C_{++}-C_{--} ,
\end{eqnarray}
where $C_{0}$ is the correlation between the occupancy of two adjacent sites
\begin{eqnarray}
&C_{0}=P\left(r,r+1\right)-P(r)P(r+1) ,
\end{eqnarray}
and $C_{\sigma,\sigma'}$ is the correlation between the occupancy of two adjacent sites whose last step was in the $\sigma$ and $\sigma'$ directions
\begin{eqnarray}
C_{\sigma,\sigma'}=P\left(r,\sigma;r+1,\sigma'\right)-P\left(r,\sigma\right)P\left(r',\sigma\right) .\label{c1}
\end{eqnarray}

The derivation is the same even for an infinite memory under one condition. In an infinite memory we assume that the transition between the (infinite) states is defined by an irreducible stochastic operator ${\cal M}$. This operator has a single eigenvalue equal to $1$ and the other eigenvalues are strictly smaller than $1$ in absolute value. If 
\begin{eqnarray}
\sup_{n\neq 1}\Re\lambda_{n}<1 ,\label{sup_ineq}
\end{eqnarray}
then the above derivation follows the same steps. However, if the inequality in (\ref{sup_ineq}) is not satisfied, i.e. that the supremum is equal to $1$, then a more subtle approach is needed.

A short note regarding the totally persistent case is in order. In this case, the model may be thought of as a $2d$-species totally antisymmetric exclusion principle (TASEP) \cite{Spitzer1970}, with equal populations of particles moving in each direction. In this model, all motion stops after a short, density-dependent relaxation time, since a particle stops moving as soon as it encounters a block containing at least one other particle of the opposite species. Therefore, in the long time limit the current is zero, and the hydrodynamic approximation breaks down.

\section{Correlations}
\label{sec_corr}
In this section we investigate the correlations between the states of two sites in an infinite system at the steady state. Due to translation invariance, the correlations depend only on the distance between the two sites. We now present a sketch of the derivation of the low-density approximation of the correlations for a general $d$-dimensional model, with the full details given in Appendix \ref{app_corr}.

Similarly to the way in which the evolution equation for the one-point function $P\left(\mathbf{r},\eta\right)$ depends on two-point correlations, see Eq. (\ref{eq1}), the evolution equation for $n$-point correlation functions depends on $(n+1)$-point correlation functions. Therefore, in order to have a finite and closed set of equations for the two-point correlations, we approximate three-point correlation functions by
\begin{eqnarray}
&P\left(\mathbf{r},\eta;\mathbf{r}',\eta';\mathbf{r}''\right)\approx\frac{1}{3}\left(1-\delta_{\mathbf{r},\mathbf{r}'}\right)\left(1-\delta_{\mathbf{r},\mathbf{r}''}\right)\left(1-\delta_{\mathbf{r}',\mathbf{r}''}\right)\nonumber\\
&\left[P\left(\mathbf{r},\eta;\mathbf{r}',\eta'\right)P\left(\mathbf{r}''\right)+P\left(\mathbf{r},\eta;\mathbf{r}''\right)P\left(\mathbf{r}',\eta'\right)+P\left(\mathbf{r}',\eta';\mathbf{r}''\right)P\left(\mathbf{r},\eta\right)\right] ,
\end{eqnarray}
where the extra Kronecker delta functions are needed to keep the approximation equal to zero if two of the sites are the same.
Under this approximation, we find two methods to derive the correlations. The first method is more cumbersome, but is applicable to all dimensions, while the second one applies only to one-dimensional systems. In both methods, we define the vector $\textbf{P}_{2}\left(\mathbf{r}\right)$ whose elements are the correlations between two sites separated by $\mathbf{r}$ occupied by particles with histories $\eta$ and $\eta'$. 

In the first method we find that the correlation functions for general $\mathbf{r}$ depend on the correlations for adjacent sites by
\begin{eqnarray}
\textbf{P}_{2}\left(\mathbf{r}\right)=\sum_{\sigma'\mathbf{d}'}{\cal Q}_{\mathbf{r},\sigma'\mathbf{d}'}\textbf{P}_{2}\left(\sigma'\mathbf{d}'\right) ,
\end{eqnarray}
where ${\cal Q}_{\mathbf{r},\sigma'\mathbf{d}'}$ is a matrix which itself depends on the memory matrix. Setting $\mathbf{r}=\sigma\mathbf{d}$ we have a closed set of linear equations between the $2d$ vectors $\textbf{P}_{2}\left(\sigma\mathbf{d}\right)$, which may be written as $\textbf{P}_{2}={\cal N}\textbf{P}_{2}$. Hence, $\textbf{P}_{2}\left(\sigma\mathbf{d}\right)$ is found by finding the unit eigenvalue of the matrix ${\cal N}$. Since the eigenvector is found up to a multiplicative constant, we use another boundary condition that at $\left|\mathbf{r}\right|\rightarrow\infty$ the two sites are uncorrelated and thus the elements of $\textbf{P}_{2}(\infty)$ are given by $P_{ss}(\eta)P_{ss}(\eta')$ where $P_{ss}(\eta)$ is the probability that a particle with history $\eta$ is in the steady state.

In the second method, which is applicable only in one-dimensional systems, we find that $\textbf{P}_{2}(r)$ satisfies
\begin{eqnarray}
\left(\begin{array}{c}\textbf{P}_{2}(r+1)\\\textbf{P}_{2}(r)\end{array}\right)={\cal Q}^{r-1}{\cal Q}_{0}\left(\begin{array}{c}\textbf{P}_{2}(1)\\0\end{array}\right) ,
\end{eqnarray}
where the matrices ${\cal Q}$ and ${\cal Q}_{0}$ are much simpler than in the first method. Again, the normalisation is taken from the requirement that at $r\rightarrow\infty$ the correlations decay to zero and the vector $\textbf{P}_{2}(\infty)$ is equal to the steady state distribution of two uncorrelated sites.

Taking for example the simplest case, a one-dimensional system with one-step memory, we find after straightforward but cumbersome calculations that
\begin{eqnarray}
\textbf{P}_{2}(r)=\frac{\rho^{2}}{4}\left(\begin{array}{c}1\\1\\1\\1\end{array}\right)+x^{r}_{1}\textbf{X}_{1}+x^{r}_{2}\textbf{X}_{2} ,\label{corr11_app}
\end{eqnarray}
where $\textbf{X}_{1}$ and $\textbf{X}_{2}$ are vectors whose exact dependence on $\rho$ and $\delta$ is too cumbersome to write explicitly, and $x_{1}$ and $x_{2}$ are
\begin{eqnarray}
\fl x_{1}=\frac{6-2\delta\rho-\sqrt{3\left(1-2\delta\right)\left(9+6\delta-4\delta\rho\right)}}{3+6\delta-2\delta\rho} ,\nonumber\\
\fl x_{2}=\frac{9+18\delta-24\delta\rho+4\delta\rho^{2}-\sqrt{3\left(3+6\delta-4\delta\rho\right)\left(9+18\delta-36\delta\rho+8\delta\rho^{2}\right)}}{4\delta\rho\left(3-\rho\right)} .
\end{eqnarray}
For small densities, we expand $\textbf{P}_{2}(r)$ to second order in $\rho$ and find that the two-point correlations are
\begin{eqnarray}
\fl C_{+,+}(r)=C_{-,-}(r)=\frac{1}{2}\left[C_{+,-}(r)+C_{-,+}(r)\right]=\frac{\rho^{2}}{4}\frac{1-8\delta-4\delta^{2}+\left(1+2\delta\right)^{2}x_{0}}{\left(1-2\delta\right)\left(1-4\delta-4\delta^{2}\right)}x^{r}_{0} ,\nonumber\\
\fl\frac{1}{2}\left[C_{+,-}(r)-C_{-,+}(r)\right]=\frac{\rho^{2}}{2}\frac{-1+x_{0}}{1-4\delta-4\delta^{2}}x^{r}_{0} ,
\end{eqnarray}
with
\begin{eqnarray}
x_{0}=\frac{2-\sqrt{4-\left(1+2\delta\right)^{2}}}{1+2\delta} .
\end{eqnarray}

\section{Comparison to numerical results}
\label{sec_numerical}
In this section we compare our analytical derivations to the numerical simulations in a one-dimensional system with one-step memory. We start from the correlations, since the analysis of the effective diffusion equation depends on them.

\subsection{Correlations}
We simulated a one-dimensional system with one-step memory in a periodic lattice. We set at $t=0$ the system to be uncorrelated, and let it evolve. After a relatively short transient time the correlation converge to their steady state values. We consider two site correlations of the form
\begin{eqnarray}
C_{\sigma,\sigma'}(r,r')=P\left(r,\sigma;r',\sigma'\right)-\frac{\rho^{2}}{4} .\label{c2}
\end{eqnarray}
\textcolor{black}{Equation (\ref{c2}) is the same as Eq. (\ref{c1}), where we note that in the steady state in a periodic lattice all sites have the same probability of being occupied, $\rho$, and the two types of memory ($\sigma=\pm$) have the same probability, $1/2$.}
Due to translation invariance, the correlations only depend on the distance between the sites, and we write $C_{\sigma,\sigma'}(r)\equiv C_{\sigma,\sigma'}(r+r',r')$. Due to inversion symmetry, we may assume that $r>0$, and further note that $C_{+,+}(r)=C_{-,-}(r)$ since both types of correlations consider two particles moving in the same direction. 

\begin{figure}
\centering
\includegraphics[width=0.3\columnwidth]{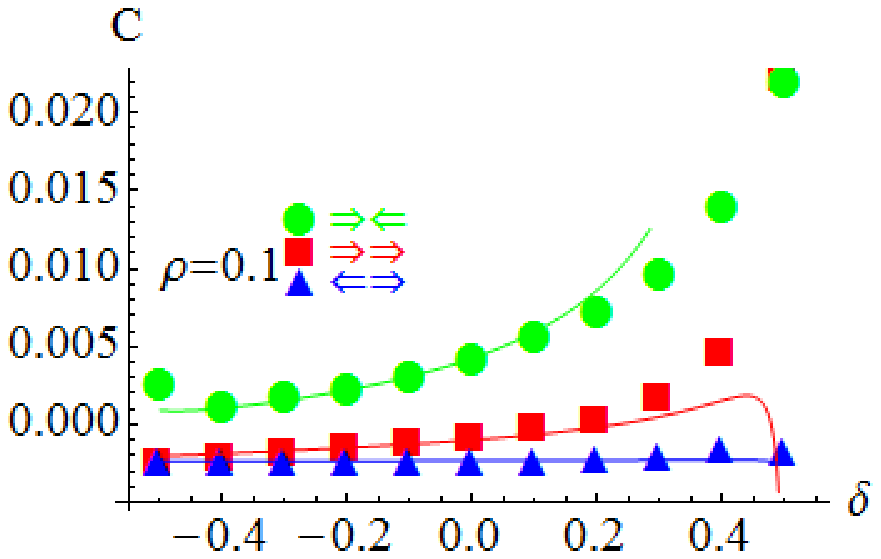}
\includegraphics[width=0.3\columnwidth]{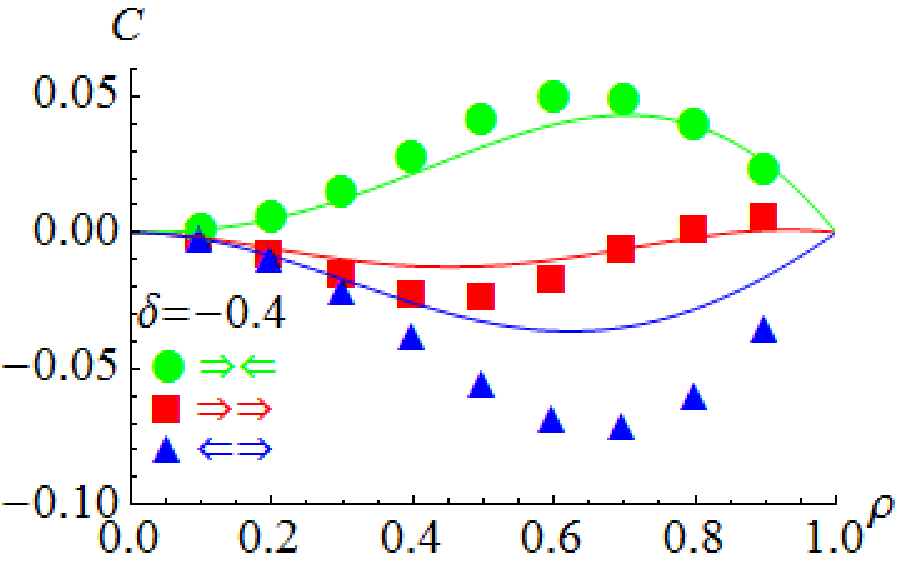}
\includegraphics[width=0.3\columnwidth]{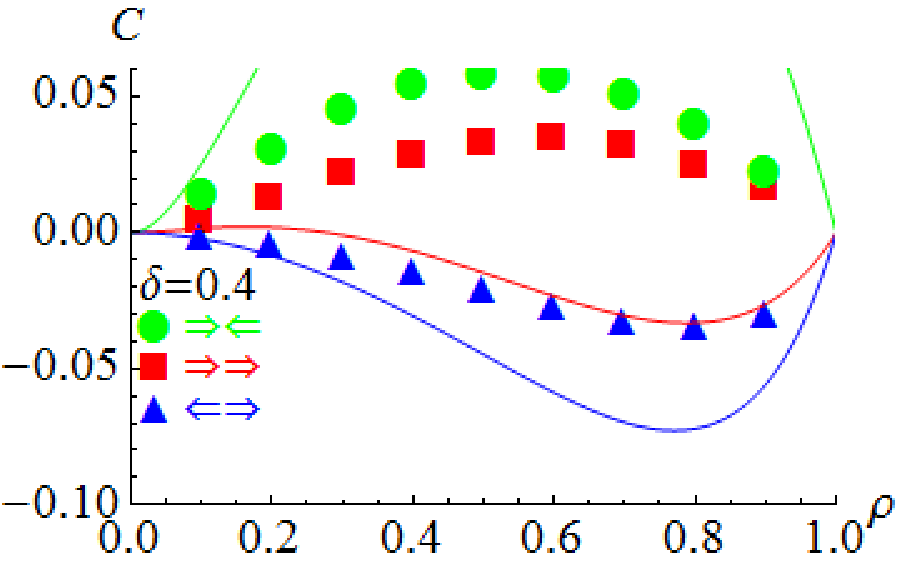}
\caption{Correlations between neighbouring sites from simulations (symbols) and the low density approximation Eq. (\ref{corr11_app}) (continuous lines) for $\rho=0.1$ (a), $\delta=-0.4$ (b) and $\delta=0.4$ (c). Each symbol represents the correlations between particles moving towards each other (green circles), away from each other (blue triangles) or in the same direction (red squares).}
\label{corr_compare}
\end{figure}

First, we compare our analytical approximation to the numerical results, and generally find that it is valid for low densities and high anti-persistent $\delta<0$, but breaks down at high persistence $\delta>0$ and high densities. Specifically, we see from figure \ref{corr_compare} that the approximation is better for the correlations between particles moving towards each other, but less so for particles moving away from each other or in the same direction. For particles moving in the same direction, the approximation might not even be qualitatively correct and get the opposite sign of the correlations for high $\delta$. One of the main conclusions from the disagreement between the numerical results and the analytical approximation, is that \textit{multi-particle correlations} are important.

\begin{figure}
\centering
\includegraphics[width=0.4\columnwidth]{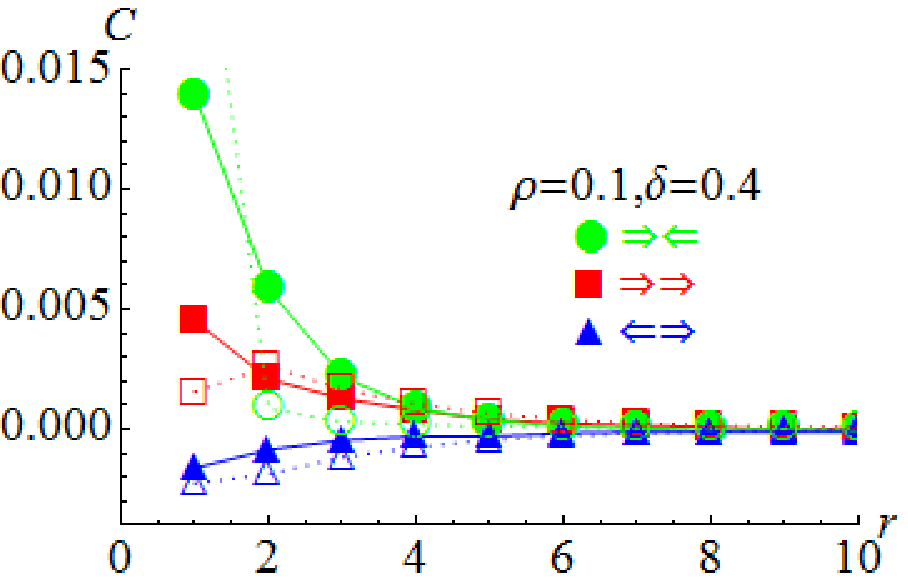}
\includegraphics[width=0.4\columnwidth]{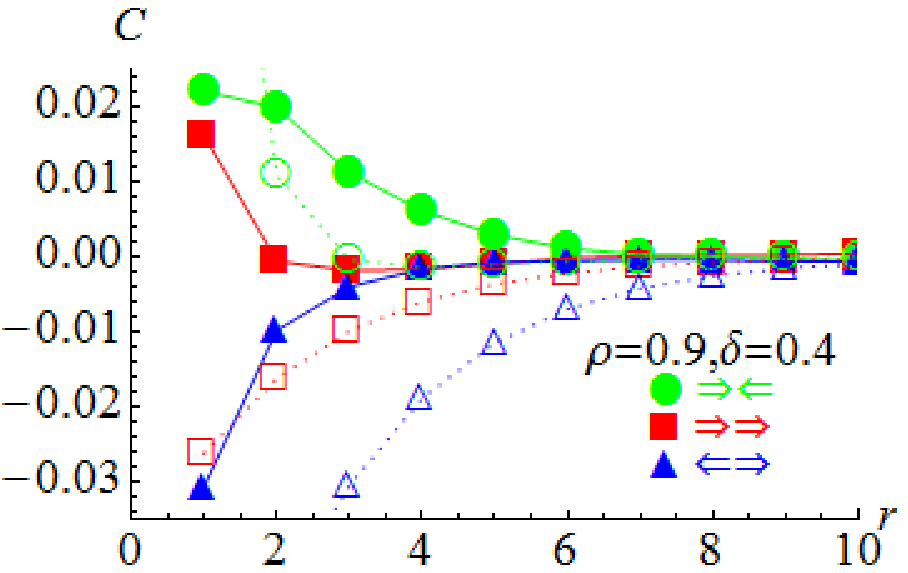}\\
\includegraphics[width=0.4\columnwidth]{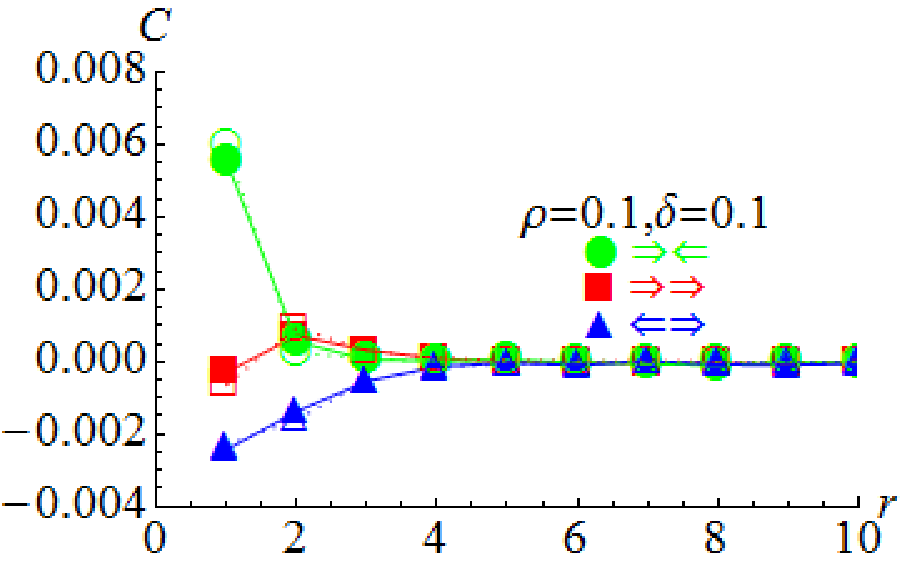}
\includegraphics[width=0.4\columnwidth]{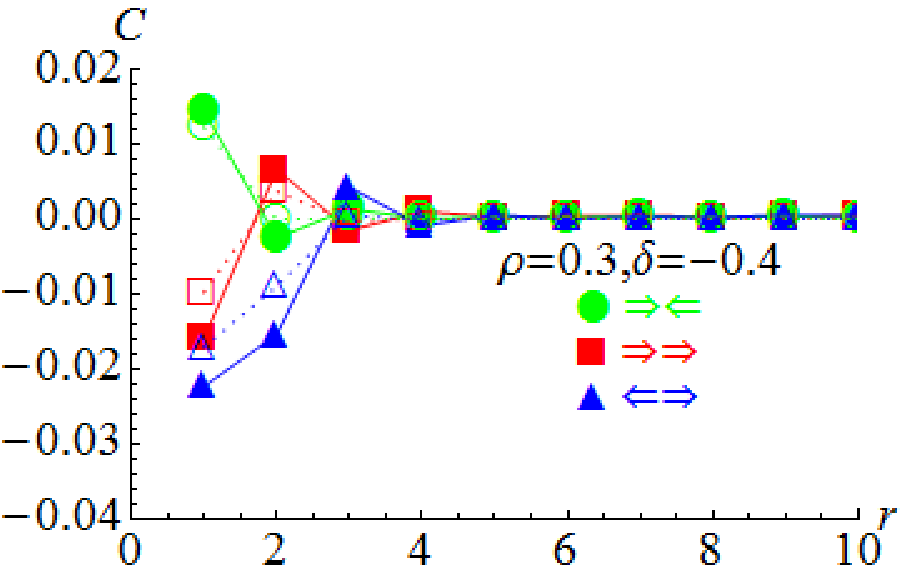}
\caption{Representative plots of the correlations $C$ as a function of the distance $r$ for various $\delta$ and $\rho$. Each symbol represents correlations between particles moving towards each other (green circles), away from each other (blue triangles) or in the same direction (red squares). The full symbols connected by continuous lines are simulation results and the empty symbols connected by dotted lines are the analytical approximation, Eq. (\ref{corr11_app}). The lines are a guide to the eye.}
\label{corr_fig}
\end{figure}

\begin{figure}
\centering
\includegraphics[width=0.4\columnwidth]{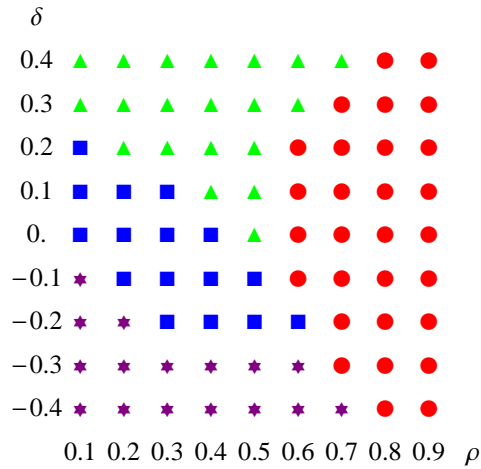}
\caption{Phase diagram of the correlations in the $\delta$-$\rho$ plane. Each symbol represents one of the four general behaviours: exponential decay (green triangle), a minimum in $C_{+,+}$ (red circle), a maximum in $C_{+,+}$ (blue square), and oscillating behaviour (purple stars).}
\label{corr_phase}
\end{figure}

Figure \ref{corr_fig} shows the four typical behaviours of the correlations. In the first type, shown in \ref{corr_fig}a, the three correlations decay to zero. In the second type, shown in \ref{corr_fig}b, the correlations between particles moving in opposite directions $C_{+,-}$ and $C_{-,+}$ decay, while the correlation between particles moving in the same direction $C_{+,+}$ have a minimum at some distance. \textcolor{black}{The reason for this depletion zone, which occurs at high densities also for memory-less walkers, is that clusters are held together by particles moving towards each other, while aligned walkers do not contribute.}. In the third behaviour, the correlations between particles moving in opposite directions $C_{+,-}$ and $C_{-,+}$ decay exponentially, while the correlation between particles moving in the same direction $C_{+,+}$ have a maximum at some distance. In the fourth behaviour, shown in \ref{corr_fig}d, all three correlations oscillate. Figure \ref{corr_phase} shows the phase diagram in the $\delta-\rho$ plane.

\begin{figure}
\centering
\includegraphics[width=0.3\columnwidth]{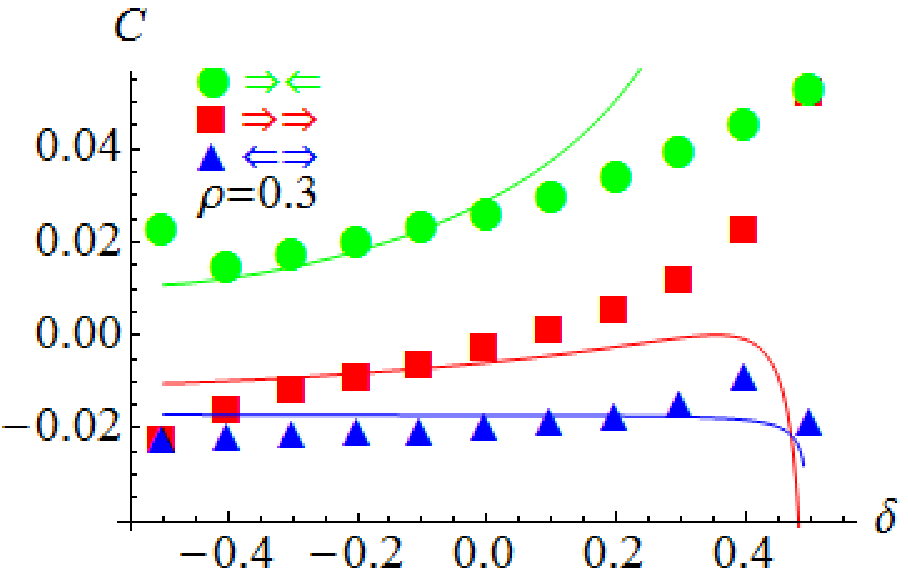}
\includegraphics[width=0.3\columnwidth]{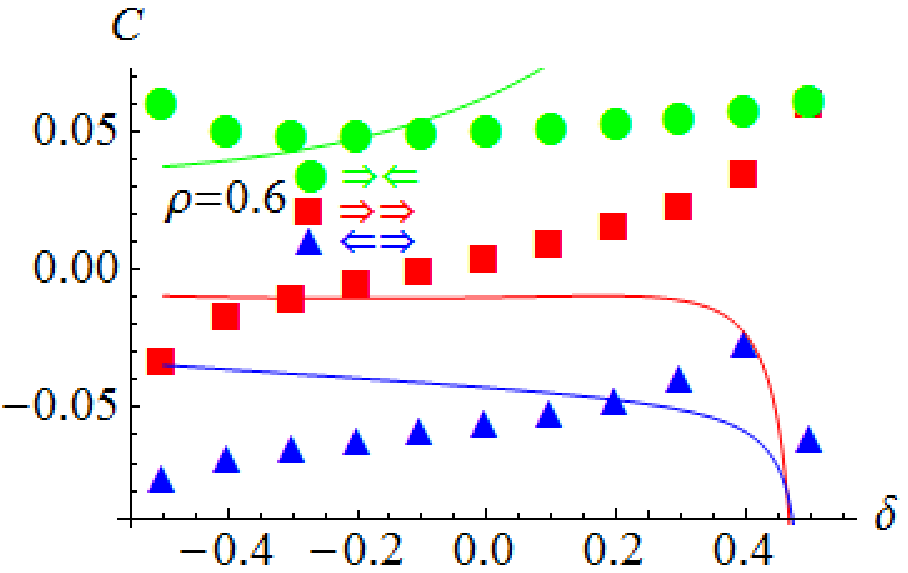}
\includegraphics[width=0.3\columnwidth]{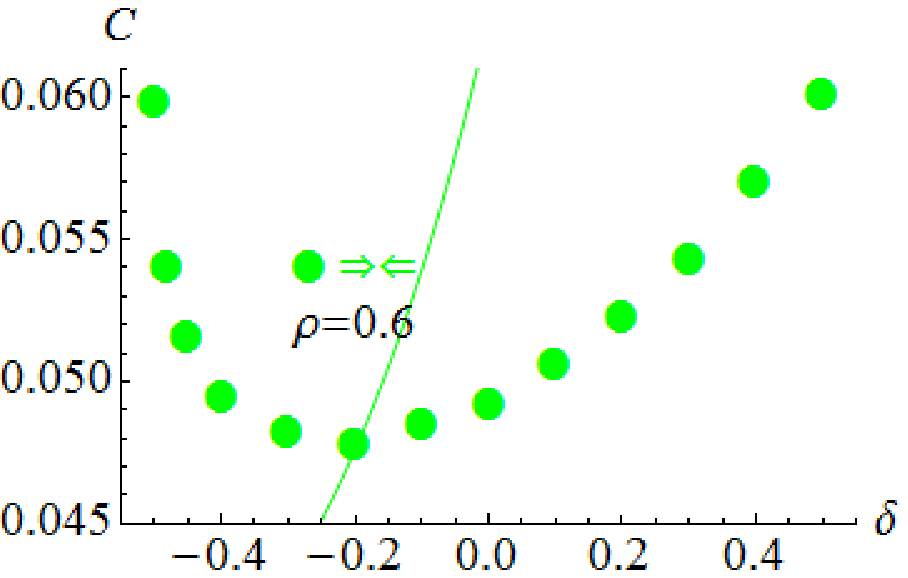}
\caption{The correlations between adjacent sites as a function of the persistence $\delta$ for $\rho=0.3$ and $\rho=0.6$. The last panel is a zoom in on the second panel. The continuous lines are the analytical approximation, and each symbol represents correlations between particles moving towards each other (green circles), away from each other (blue triangles) or in the same direction (red squares).}
\label{corr_vs_d}
\end{figure}

We are especially interested in the correlations between nearest neighbours, $C_{\sigma,\sigma'}(1)$. We first observe their dependence on $\delta$, as shown in figure \ref{corr_vs_d}. The correlations of particles moving in the same direction $C_{+,+}(1)$ and moving away from each other $C_{+,-}(1)$ are always increasing functions of $\delta$, except for a singularity of $C_{+,-}(1)$ at $\delta=\frac{1}{2}$. The correlation between particles moving towards each other $C_{-,+}$ is an increasing function of $\delta$ for $\rho\leq1/2$, while for $\rho>1/2$ it is non-monotonic with a single minimum. \textcolor{black}{The minimum in $C_{-,+}$ can be explained as follows. For $\delta\approx\frac{1}{2}$ the correlation is always increasing with $\delta$ since adjacent particles are more likely to remain so. For $\delta\approx-\frac{1}{2}$ and $\rho>\frac{1}{2}$, consider two particles at the edge of a cluster which in the last step tried to move toward each other. Assuming the particle at the edge moves away, then the closer $\delta$ is to $-\frac{1}{2}$, the more likely it is to return to the edge of the cluster at the next step, and thus the correlation $C_{-,+}$ is higher. Therefore, $C_{-,+}$ decreases with $\delta$ at $\delta\approx-\frac{1}{2}$.}

\begin{figure}
\centering
\includegraphics[width=0.4\columnwidth]{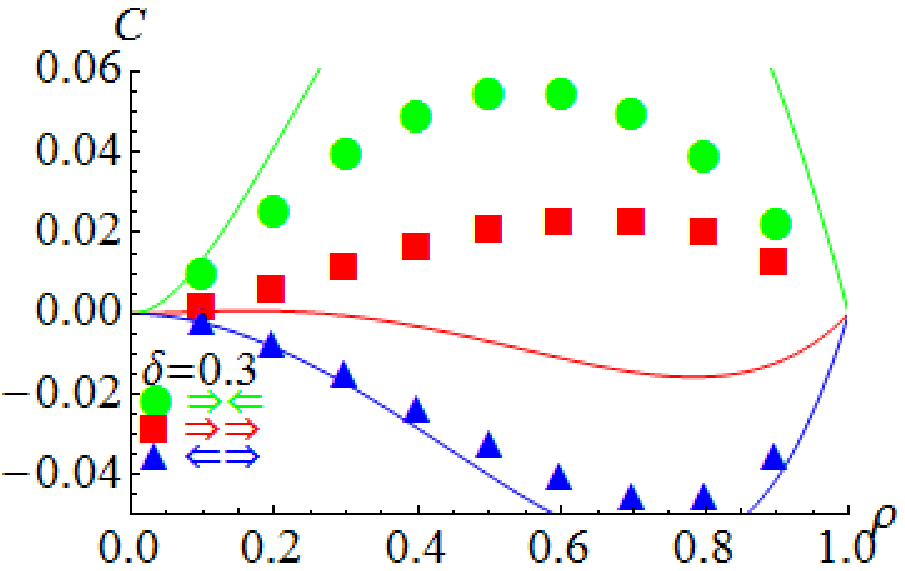}
\includegraphics[width=0.4\columnwidth]{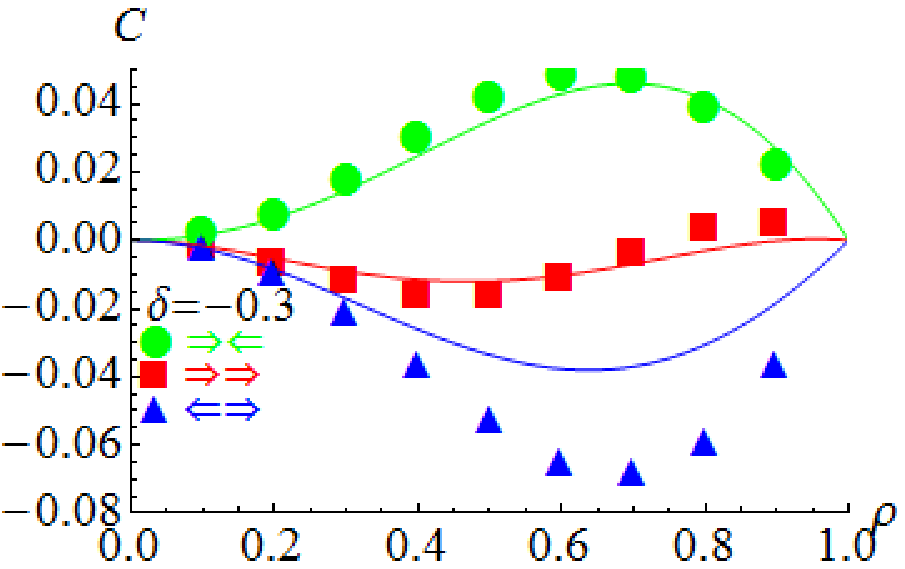}
\caption{Correlations between adjacent sites as function of the density $\rho$ for $\delta=0.3$ and $\delta=-0.3$. The continuous lines are the analytical approximation Eq. (\ref{corr11_app}), and each symbol represents correlations between particles moving towards each other (green circles), away from each other (blue triangles) or in the same direction (red squares).}
\label{corr_vs_r}
\end{figure}

We now investigate the dependence of $C_{\sigma,\sigma'}(1)$ on the density shown in figure \ref{corr_vs_r}. Note that all correlations should vanish at $\rho=0$ and $\rho=1$. We find that the correlation between particle moving towards each other $C_{-,+}$ has a maximum, while the correlation between particles moving away from each other $C_{+,-}$ has a minimum. The behaviour of the correlation between particles moving in the same direction depends on whether $\delta$ is positive or negative. For $\delta>0$, we find that $C_{+,+}$ has a maximum, while for $\delta<0$ it has a minimum.

\begin{figure}
\centering
\includegraphics[width=0.4\columnwidth]{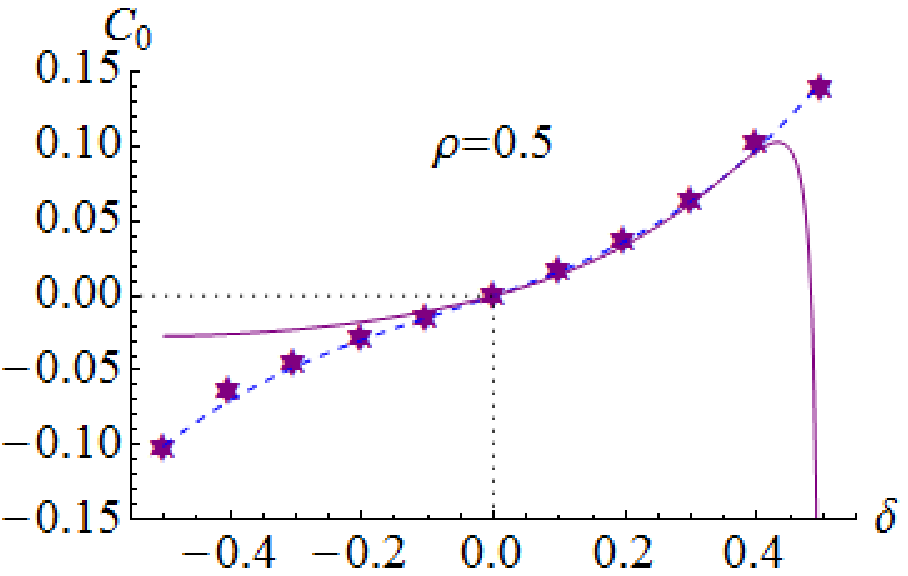}
\includegraphics[width=0.4\columnwidth]{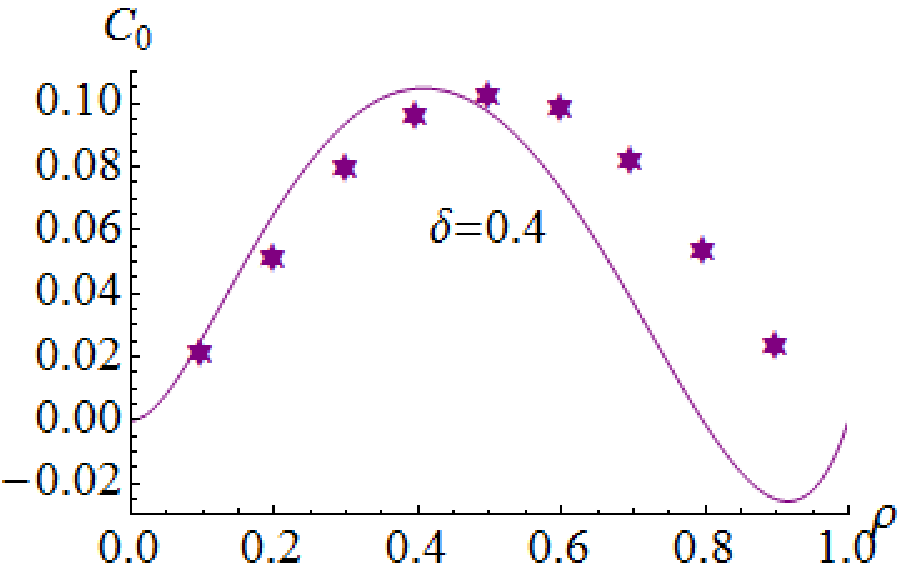}
\caption{Dependence of the correlation $C_{0}(1)$ on $\delta$ for $\rho=0.5$ (a) and on $\rho$ for $\delta=0.4$ (b). It is qualitatively similar for all other parameters. The continuous purple line is the analytical approximation Eq. (\ref{corr11_app}), and the dashed blue line in panel (a) is a fit to a cubic polynomial Eq. (\ref{c0_cube}). The dotted line in panel (a) shows that the correlation vanishes at $\delta=0$.}
\label{corr0_dr}
\end{figure}

\begin{figure}
\centering
\includegraphics[width=0.4\columnwidth]{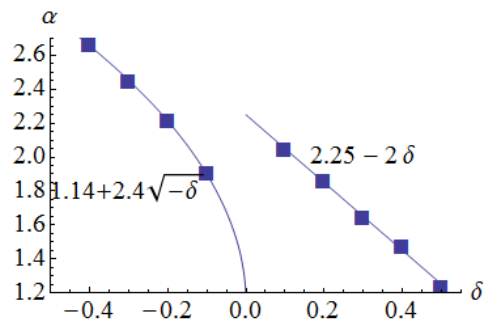}
\caption{Dependence of the exponent $\alpha$ on the persistence $\delta$. The continuous lines are fits, see equation (\ref{alpha_def}) for the definition.}
\label{corr_aexp}
\end{figure}

Next, we consider the correlations between the occupancy of different \textcolor{black}{sites} $C_{0}(r)$ regardless of their history
\begin{eqnarray}
C_{0}(r)=2C_{+,+}(r)+C_{+,-}(r)+C_{-,+}(r) .
\end{eqnarray}
We find that this correlation decays exponentially with $r$. It is positive for $\delta>0$ and negative for $\delta<0$. The dependence of $C_{0}(1)$ on $\delta$ and $\rho$ is shown in figure \ref{corr0_dr}. We find \textcolor{black}{numerically} that the dependence on the density $\rho$ is symmetric around $\rho=1/2$ and captured by
\begin{eqnarray}
C_{0}(r=1,\rho,\delta)=C_{0}\left(r=1,\rho=\frac{1}{2},\delta\right)\left[4\rho\left(1-\rho\right)\right]^{\alpha} ,\label{alpha_def}
\end{eqnarray}
with the dependence of the exponent $\alpha$ on the persistence $\delta$ shown in figure \ref{corr_aexp}. Fitting the correlations at $\rho=\frac{1}{2}$ to a cubic polynomial that vanishes at $\delta=0$ yields
\begin{eqnarray}
C_{0}\left(r=1,\rho=\frac{1}{2},\delta\right)\approx 0.15\delta + 0.089\delta^2 + 0.38\delta^3 .\label{c0_cube}
\end{eqnarray}

\begin{figure}
\centering
\includegraphics[width=0.4\columnwidth]{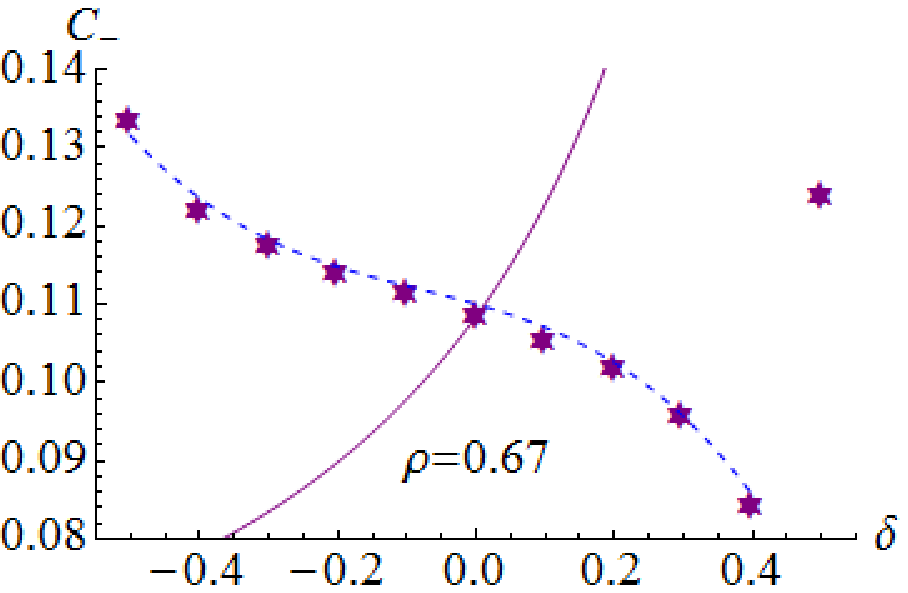}
\includegraphics[width=0.4\columnwidth]{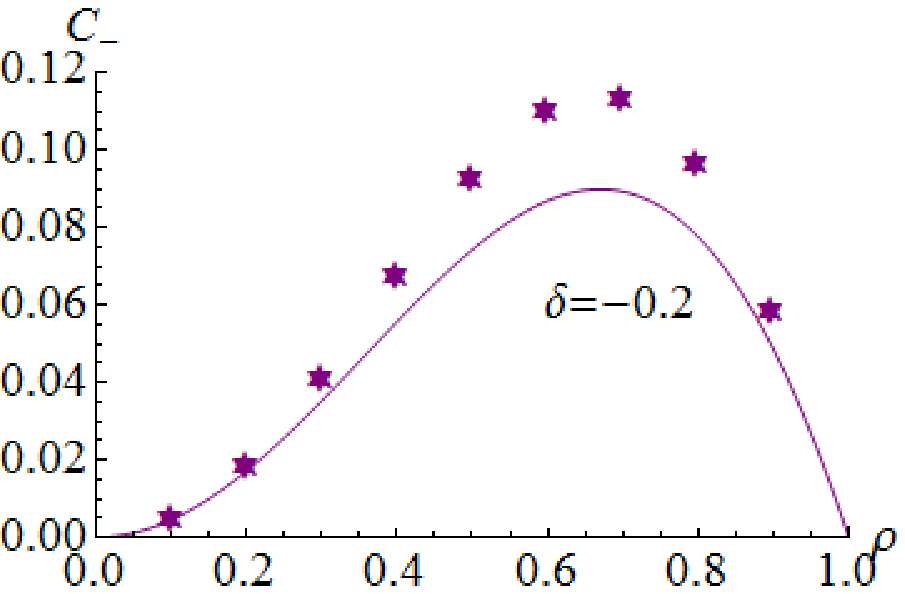}
\caption{Dependence of the correlation $C_{m}(1)$ on $\delta$ for $\rho=0.67$ (a) and on $\rho$ for $\delta=-0.2$. The continuous purple lines are the analytical approximation Eq. (\ref{corr11_app}), and the dashed blue line in panel (a) is a fit to a cubic polynomial Eq. (\ref{cm_cube}).}
\label{corr_m_dr}
\end{figure}

\begin{figure}
\centering
\includegraphics[width=0.4\columnwidth]{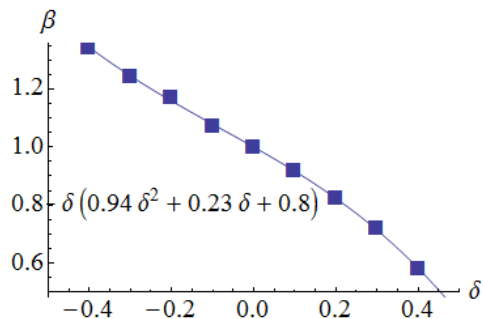}
\caption{Dependence of the exponent $\beta$ on the persistence $\delta$. The continuous line is a fit to a cubic polynomial, see equation (\ref{beta_def}) for the definition.}
\label{bexp}
\end{figure}

Another interesting correlation is $C_{-}(r)=C_{-,+}(r)-C_{+,-}(r)$ which encodes the asymmetry between the histories. This correlation decays exponentially and is almost always positive, except for the case of strong antipersistence, where it decays and oscillates. Figure \ref{corr_m_dr} shows the dependence of $C_{-}(1)$ on $\delta$ and $\rho$. We find that it is well described by
\begin{eqnarray}
&C_{-}\left(r=1,\rho,\delta\right)=C_{-}\left(r=1,\rho=\frac{2}{3},\delta\right)\left[\frac{27}{4}\rho^{2}\left(1-\rho\right)\right]^{\beta} ,\label{beta_def}
\end{eqnarray}
with the dependence of the exponent $\beta$ on the persistence $\delta$ shown in figure \ref{bexp}. Fitting the correlations at $\rho=2/3$ to a cubic polynomial yields
\begin{eqnarray}
&C_{-}\left(r=1,\rho=\frac{2}{3},\delta\right)\approx0.11-0.025\delta-0.034\delta^{2}-0.14\delta^{3} .\label{cm_cube}
\end{eqnarray}

\subsection{Transport}
To obtain insight into the transport behaviour of our system, we perform simulations on a one-dimensional interval of length $L$ connected to particle reservoirs at sites $r=0$ and $r=L+1$ with densities $\rho_{0}$ and $\rho_{L}$ respectively. The initial condition inside the system is such that for sites $1$ to $L$ the probability to be occupied at time $0$ is a linear interpolation between $\rho_{0}$ and $\rho_{L}$. After a relatively short transient time the mean density converges to a steady state.

\begin{figure}
\centering
\includegraphics[width=0.45\columnwidth]{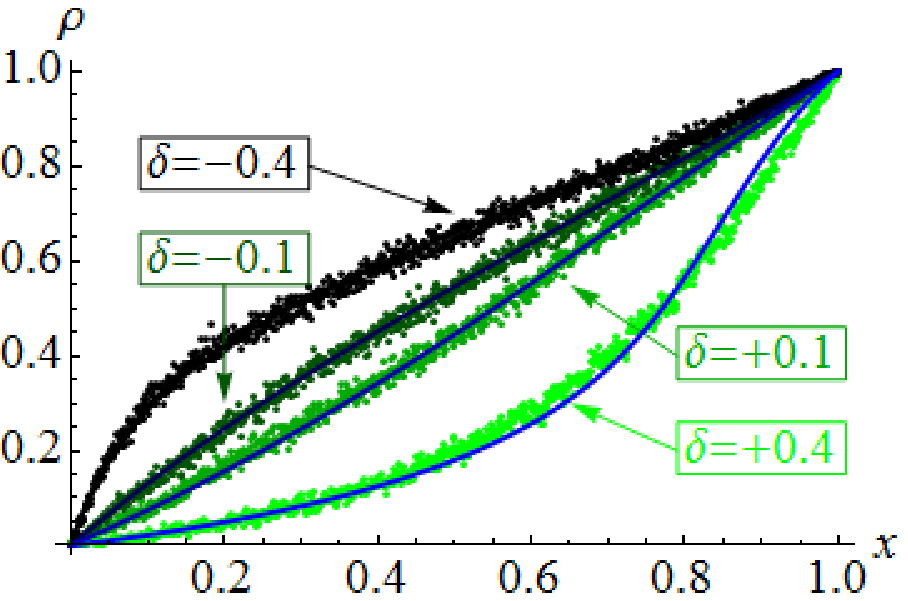}
\includegraphics[width=0.45\columnwidth]{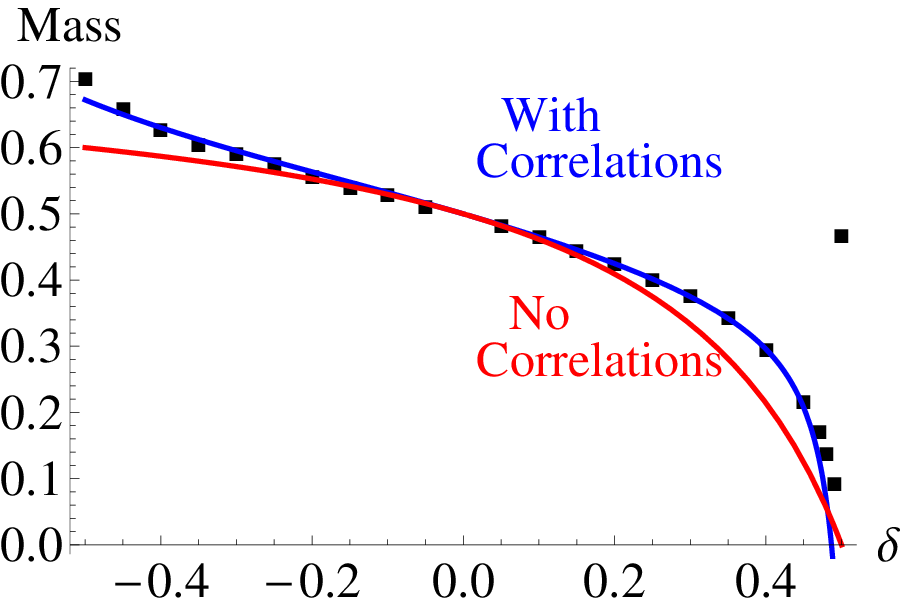}
\caption{Steady state profile (a) and the total mass in the system (b) at the steady state for different values of the persistence $\delta$. The continuous lines are the analytical results, Eqs. (\ref{steady_gen}) and (\ref{mass_gen}), with correlations taken from the numerical fits, not the analytical approximation.}
\label{1d_transport_fig}
\end{figure}

For a known diffusion coefficient, the steady state profile is given by \cite{Teomy2017}
\begin{eqnarray}
\frac{x}{L}=\frac{\int^{\rho(x)}_{\rho_{0}}D\left(\rho'\right)d\rho'}{\int^{\rho_{L}}_{\rho_{0}}D\left(\rho'\right)d\rho'} .\label{steady_gen}
\end{eqnarray}
\textcolor{black}{For any given $D(\rho)$,} by evaluating the integral and inverting the equation, we \textcolor{black}{can} find an analytical expression for the steady state. Figure \ref{1d_transport_fig}(a) shows the steady state profile for the one-step memory model with reservoir densities $\rho_{0}=0$ and $\rho_{0}=1$. We see a very good agreement between the numerical simulations, and the semi-analytical result based on Eq. (\ref{steady_gen}) when the correlations are taken from the non-biased simulations. In figure \ref{1d_transport_fig}(b) we show the total mass in the steady state, given by \cite{Teomy2017}
\begin{eqnarray}
M=\frac{1}{L}\int^{L}_{0}\rho(x)dx=\frac{\int^{\rho_{L}}_{0}\rho D(\rho)d\rho}{\int^{\rho_{L}}_{\rho_{0}}D(\rho)d\rho} ,\label{mass_gen}
\end{eqnarray}
and again we find good agreement between the numerical simulations and the semi-analytical expression \textcolor{black}{when the correlations are taken into account. When the correlations are neglected, we find that the estimated mass is always lower than in the simulations. Note that at $\delta=0$ the relevant correlations are indeed equal to zero, and therefore for small values of $\delta$, neglecting the correlations is justified.}

\section{Full anti-persistence}
\label{sec_tap}
In the extreme limit of full anti-persistence the system exhibits several unique properties. We consider a one-dimensional lattice with totally anti-persistent particles, such that at each step the particles always switch direction and attempt to move in the opposite direction than before. In a closed system with density $\rho<1/2$, \textcolor{black}{the absorbing states of the system are such that each particle jiggles between two sites, and thus at the steady state the system relaxes to them.} If the density is higher than $1/2$ or the system is open, the situation is different. A special case is $\rho=1/2$ in a closed system. To our knowledge, this pathological case has not been explored before. The two sites between which the particle hops change only if another particle enters one of the two sites, such that the first walker pushes itself on its new neighbour. Physically, this limit \textcolor{black}{may represent two different scenarios. First, it may represent a series of very deep and narrow traps, such that a lone particle cannot escape (qualitatively similar to the way a single particle hops between two sites), but if another particle enters the same trap, one of them must leave. Secondly, it may represent the movement of motors in a highly viscous medium where inertia is negligible and motion is dependent upon pushing other objects.}

\begin{figure}
\centering
\includegraphics[height=4cm]{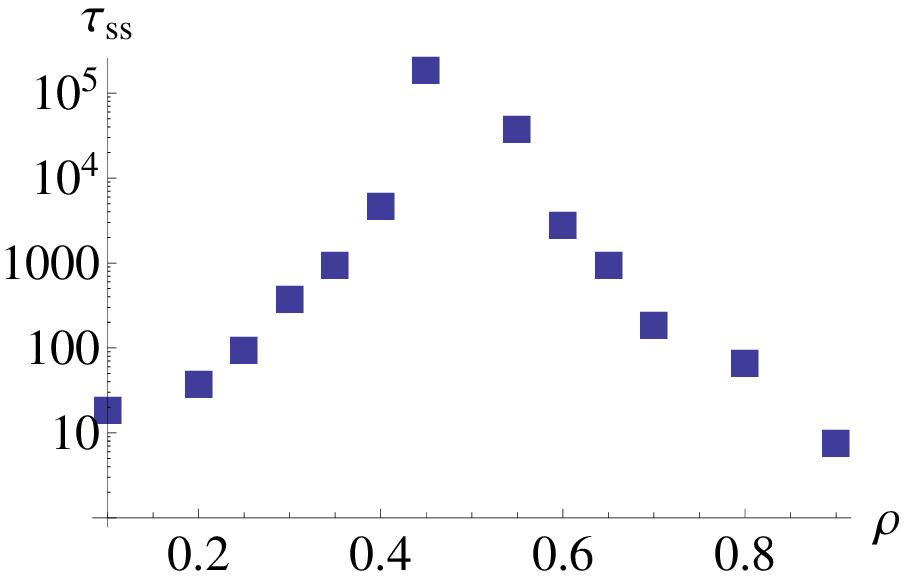}
\includegraphics[height=4cm]{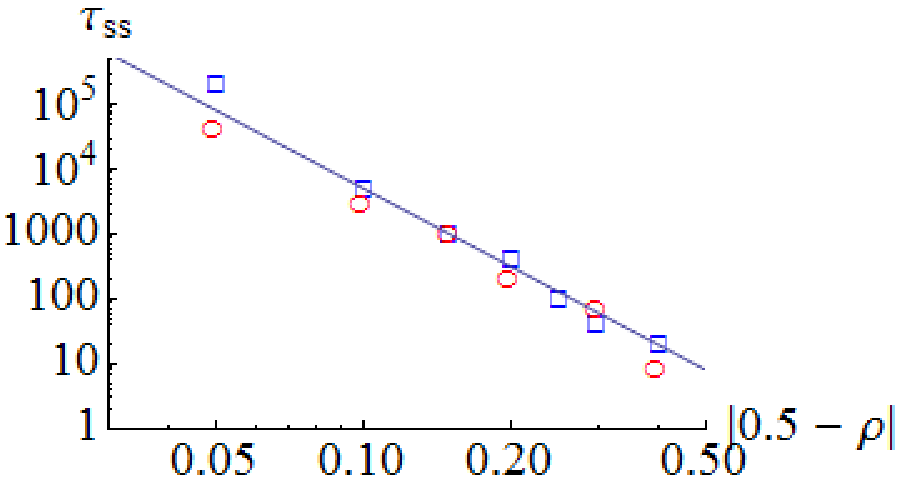}
\caption{The relaxation time $\tau_{ss}$ for the system to reach the steady state vs. the density \textcolor{black}{in log scale (a) and log-log scale (b)}. It diverges at $\rho=1/2$. \textcolor{black}{In panel (b) blue squares are the data for $\rho<1/2$ and the red circles are the data for $\rho>1/2$. The straight line is $|\frac{1}{2}-\rho|^{-4}$.}}
\label{tap_corr_time}
\end{figure}

\begin{figure}
\centering
\includegraphics[width=0.45\columnwidth]{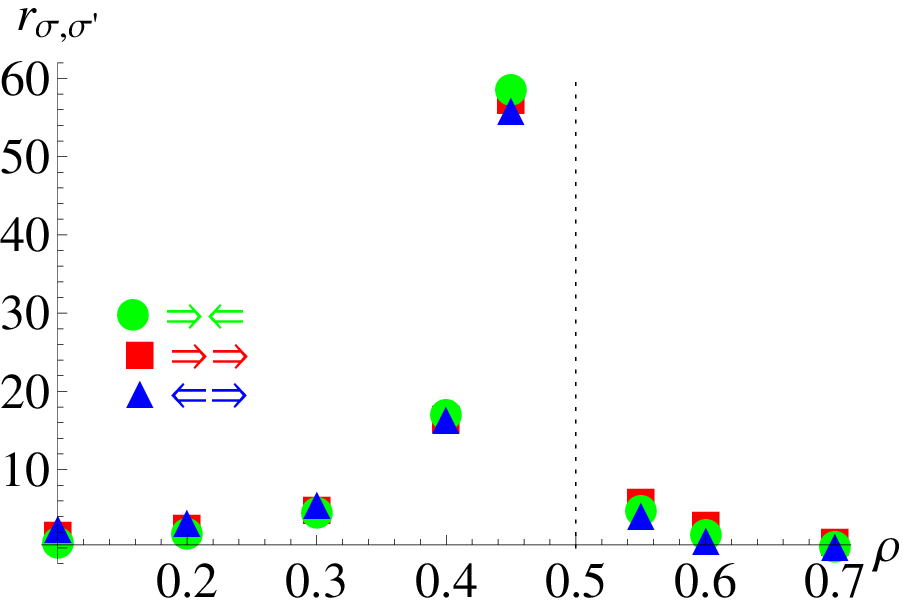}
\includegraphics[width=0.45\columnwidth]{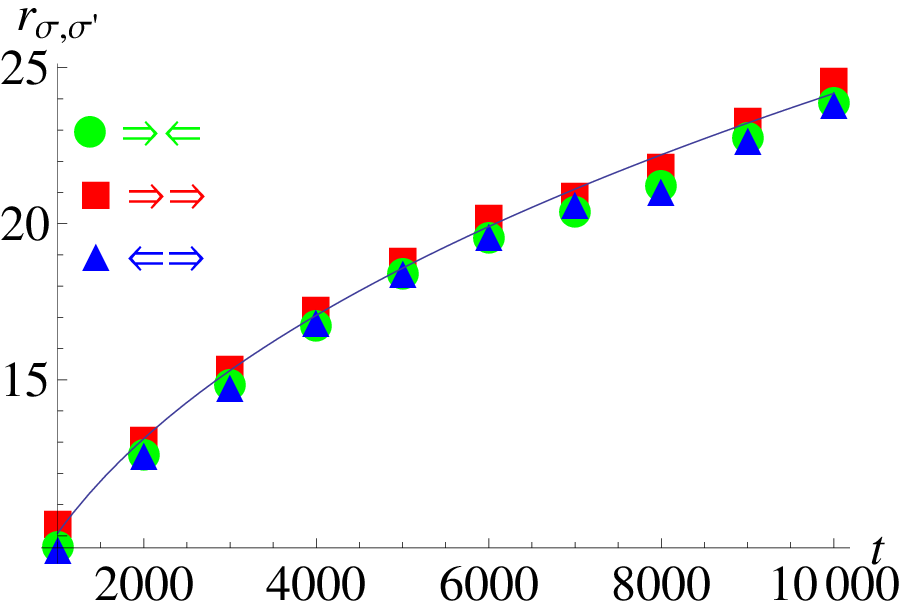}
\caption{The three possible correlation lengths vs. the density for $\rho\neq1/2$ (a) and as a function of time for $\rho=1/2$. The continuous line in panel (b) is $0.73t^{0.38}$. Each symbol represents correlations between particles moving towards each other (green circles), away from each other (blue triangles) or aligned (red squares).}
\label{tap_corr_length}
\end{figure}

First we consider the correlations in a closed system. If $\rho\neq1/2$ the correlations converge to a steady state value after some time $\tau_{ss}$ as shown in figure \ref{tap_corr_time}. We note that $\tau_{ss}$ diverges at $\rho=1/2$. In the steady state, the correlations oscillate with an exponentially decaying envelope, $e^{-r/r_{\sigma,\sigma'}}$, with $r_{\sigma,\sigma'}$ shown in figure \ref{tap_corr_length}(a). We note that the three correlation lengths $r_{+,+},r_{+,-}$ and $r_{-,+}$ are approximately the same. For low ($\rho<0.3$) and high ($\rho>0.6$) densities the correlation length is very small ($\lesssim 2$). However it diverges at $\rho=1/2$. Note that the in the low density regime ($\rho<\frac{1}{2}$) the correlation length is much higher than in the high density regime ($\rho>\frac{1}{2}$). In the special case $\rho=1/2$ the correlations do not converge to a steady state and the correlation length increases with time as shown in figure \ref{tap_corr_length}(b). We find that $r_{\sigma,\sigma'}\simeq t^{0.38}$.

\begin{figure}
\centering
\includegraphics[width=0.45\columnwidth]{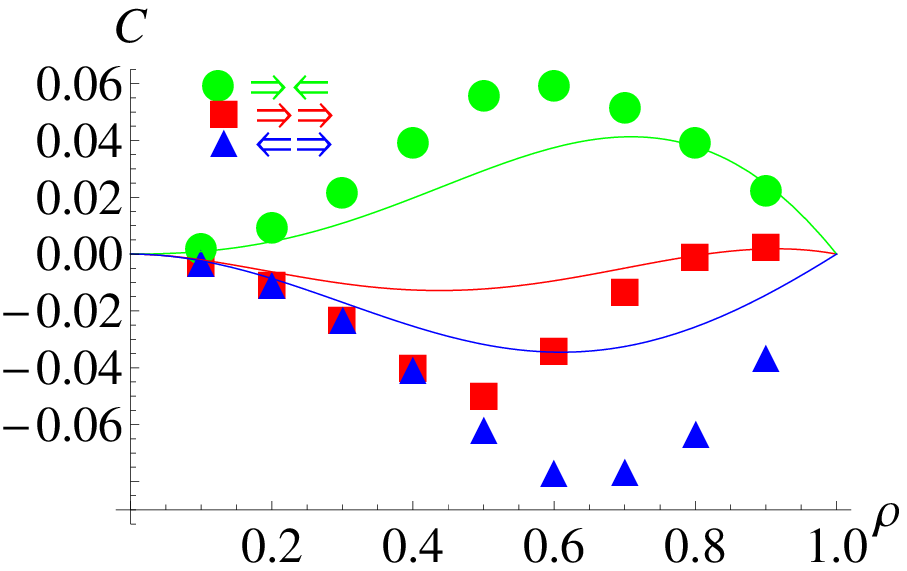}
\includegraphics[width=0.45\columnwidth]{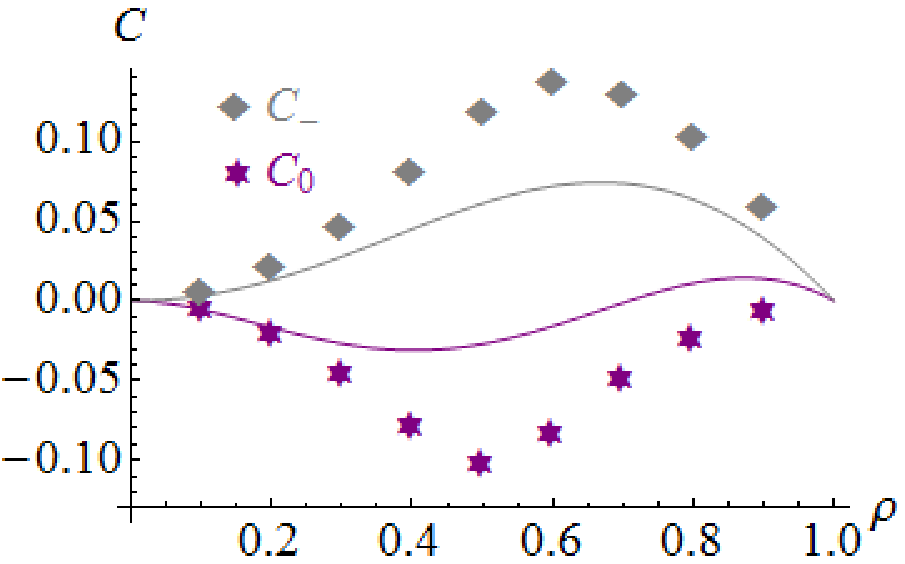}
\caption{Correlations between adjacent sites as function of the density. In (a) each symbol represents correlations between particles moving towards each other (green circles), away from each other (blue triangles), or in the same direction (red squares). \textcolor{black}{Panel (b) shows the correlations $C_{-}$ and $C_{0}$.} The continuous lines are the analytical approximation Eq. (\ref{corr11_app}).}
\label{tap_corr_r}
\end{figure}

Although the correlation length diverges at $\rho=1/2$, for any finite distance the correlations do converge after a finite time. \textcolor{black}{The reason is that locally, the particles arrange themselves into a lattice where each particle has its own two sites between which it hops.} Let us concentrate on the correlations between adjacent sites, setting $r=1$. The correlations as a function of the density are shown in figure \ref{tap_corr_r}. All the various correlations behave similarly to the regular, finite $\delta$ case, \textcolor{black}{with the same general trend qualitatively captured by the analytical approximation.}

\begin{figure}
\centering
\includegraphics[width=0.45\columnwidth]{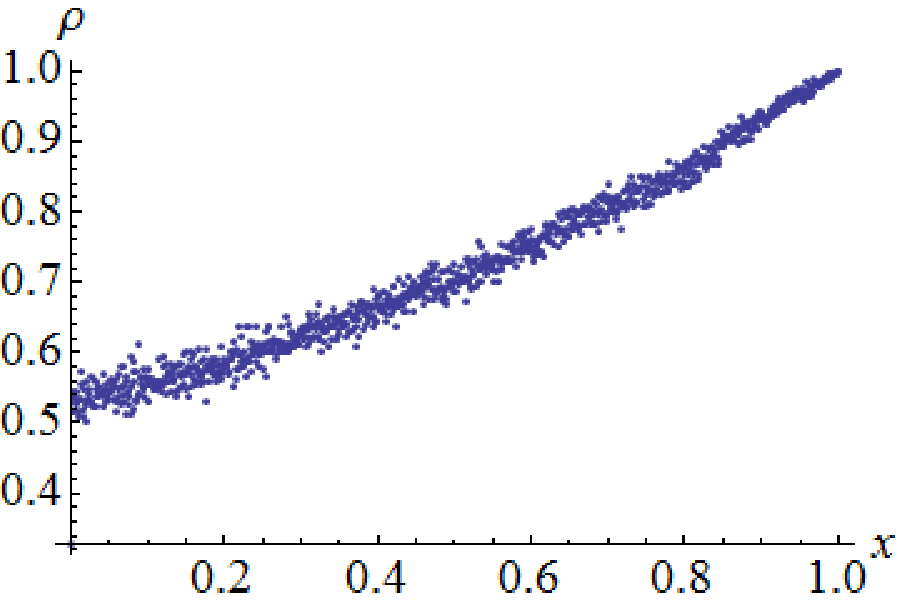}
\includegraphics[width=0.45\columnwidth]{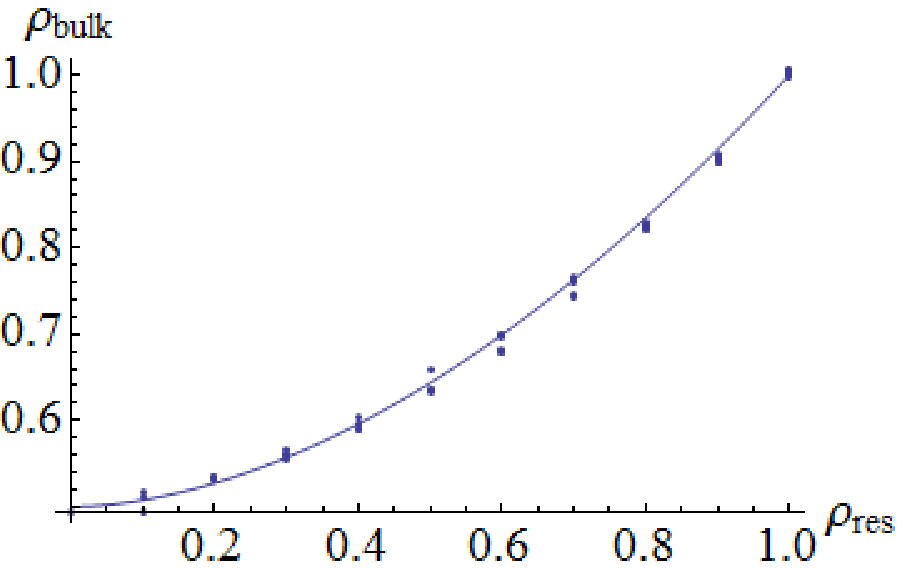}
\caption{(a) Steady state mean density profile for particle reservoirs $\rho_{0}=0.2$ and $\rho_{L}=0.7$. (b) Mean density at the system's edge vs. the reservoir density. Each symbol represents a system with different reservoir densities. The density at the system's edge depends only on the density of the reservoir near it.}
\label{tap_transport}
\end{figure}

Secondly, we now consider the transport in a system connected to reservoirs. Here, the totally-antipersistent model behaves critically different from the non-extreme case. \textcolor{black}{We find numerically, that the density profile always appears to be linear, as shown in figure \ref{tap_transport}(a). Contrast this behaviour with the highly non-linear profile for general $\delta$ shown in figure \ref{1d_transport_fig}.} As we see, the mean density at the edge of the system $\rho_{bulk}$ is different from the density of the \textcolor{black}{neighbouring} reservoir $\rho_{res}$. \textcolor{black}{The behaviour is the same at both the right and the left reservoir.} We find that $\rho_{bulk}\geq\rho_{res}$, and as shown in figure \ref{tap_transport}b it is well approximated by
\begin{eqnarray}
\rho_{bulk}=\frac{1+\rho^{1.8}_{res}}{2} .
\end{eqnarray}

In order to understand this phenomena consider an initially empty system connected to a reservoir with density $0$ and a reservoir with density $\rho_{0}$. Initially, at each time step there is a probability $\rho_{0}/2$ that the particle in the reservoir jumps to site $1$. This particle will go back to the reservoir with probability $1-\rho_{0}$, but will remain in place and change its heading with probability $\rho_{0}$. Hence, after time $2\rho^{-2}_{0}$ on average, there will be a particle that moves back and forth between sites $1$ and $2$, and thus the mean density on these sites will be $\frac{1}{2}$. This process continues, and the front eventually reaches the second reservoir. Therefore, the probability that site $1$ is occupied may be approximated as a sum of the probability that the particle moving back and forth between sites $1$ and $2$ occupies it, $\frac{1}{2}$, and the probability that it is occupied by a newly arrived particle from the reservoir $\frac{\rho^{2}_{0}}{2}$, i.e. $\rho_{bulk}\approx\frac{1+\rho^{2}_{0}}{2}$. As we see this simple argument is very close to the fitted scaling exponent.

\section{Summary}
\label{sec_summary}
We investigated a $d$ dimensional lattice gas of walkers with finite memory, in which each site is occupied by at most one particle, and the direction each particle attempts to move to depends on a specified part of its history. Specifically, we considered the two-site same time correlations and the hydrodynamic description of the general model.  

We derived a non-linear drift-diffusion equation which takes into account the correlations between the particles, and identified a memory-dependent critical density which governs the difference between the density-dependent bulk diffusion coefficient $D$ from the memory-less one $D_{0}$. For persistent walkers, below the critical density $D>D_{0}$ while above it $D<D_{0}$. For anti-persistent walkers, the situation is reversed: below the critical density $D<D_{0}$ while above it $D>D_{0}$. If the correlations are neglected the critical density is exactly $1/2$, and moreover, the diffusion coefficient has the exact same density dependence except for a single memory-dependent factor. We also derived a low-density approximation for the same time correlations between different sites, again for a general finite memory in $d$ dimensional hypercubic lattice. 

We performed simulations on a one-dimensional system with one-step memory and found excellent agreement between our analytical derivation and the numerical results. \textcolor{black}{Specifically, we considered the steady state mean density profile of a system connected to particle reservoirs at both edges. By using the correlations from the non-biased simulations we found that our analytical derivation describes the density profile very well. However, if the correlations are neglected, then the total mass in the bulk is under-estimated.}

We specifically considered the previously unexplored special case of totally anti-persistent particles. Generally, the correlations converge to their steady state values after a finite time and have a finite correlation length. However in the special case of totally anti-persistent particles and density $1/2$, the correlations do not converge and the correlation length diverges as a power law with time. 

We also studied the transport properties of an open system. To this end we connect an ensemble of totally anti-persistent particles to external particle reservoirs. In almost all systems, regardless of the precise details of the microscopic dynamics, when a system is connected to a reservoir, the mean density of particles at the edge is the same as the reservoir following the zeroth law of thermodynamics. In a totally anti-persistent system, however, the density at the edge is always higher than in the reservoir. \textcolor{black}{This is not a true breaking of the zeroth law, since the system does not obey detailed balance and is thus out of equilibrium, however to our knowledge no other model, including those describing systems out of equilibrium, behaves in a similar manner.} The explanation is that in a semi-infinite system, connected to only one reservoir, once a particle reaches the second site from the edge it will never return to the reservoir, and thus the local rate of injection from the reservoir is higher than the local rate of depletion to the reservoir. We found a simple approximation for the density at the edge of the system, which agrees reasonably well with the numerical results. A physical system which may be described by this limit is a series of very deep and narrow traps, such that each particle tends to stay in its trap, unless another particle comes and pushes it out.

Although all our derivations concern particles with finite memory, they are also valid for particles with infinite memory, as long as the velocity autocorrelations decay fast enough. If the autocorrelations decay slowly, our derivation breaks down. It would be interesting to see how our results expand to such slowly decaying correlations, even for specific realisations of the memory term. We speculate whether the resulting hydrodynamic description would be a fractional diffusion equation, embodying long-ranged memory structures \cite{Metzler2000}.

In the model we investigated here, the direction chosen at each step is completely uncorrelated to the success of failure of the moves and thus to the other particles. Another interesting expansion involves correlating the chosen direction with the density, such that the history of each particle contains also information about the success or failure of previous moves. We conjecture that persistent walkers that tend to turn around when they are blocked would exhibit a transition from a behaviour similar to positive persistence at low densities to a behaviour similar to anti-persistence at high densities.

\ack
We would like to thank Yair Shokef, Michael Urbakh and Andrey Cherstvy for fruitful discussions. ET acknowledges financial support from the TAU-Potsdam fellowship.

\appendix
\section{Derivation of the diffusion equation}
\label{app_diff}
In this section we consider the hydrodynamic description of the models and derive an effective diffusion equation. For simplicity, we first perform the derivation for the simplest case, a one-step memory in one dimension, and after that consider a general finite memory in $d$ dimensions. \textcolor{black}{For brevity, we neglect the explicit time dependence in our notation.}

\subsection{One-step memory in one dimension}
We define $P(r,\sigma)$ as the probability that site $r$ is occupied by a particle whose last step was in the $\sigma=\pm1$ direction, and $P(r,\sigma;r')$ as the probability that site $r$ is occupied by a particle whose last step was in the $\sigma$ direction and site $r'$ is occupied by a particle with any memory.
The probabilities $P(r,\sigma)$ evolve in time according to
\begin{eqnarray}
\fl\tau\frac{\partial P(r,\sigma)}{\partial t}=&-P(r,\sigma)+\left(\frac{1}{2}+\delta\right)P(r,\sigma;r+\sigma)+\left(\frac{1}{2}-\delta\right)P(r,-\sigma;r+\sigma)+\nonumber\\
&+\left(\frac{1}{2}+\delta\right)\left[P(r-\sigma,\sigma)-P(r-\sigma,\sigma;r)\right]\nonumber\\
&+\left(\frac{1}{2}-\delta\right)\left[P(r-\sigma,-\sigma)-P(r-\sigma,-\sigma;r)\right] .\label{ev1}
\end{eqnarray}
The five terms on the right hand side of (\ref{ev1}) correspond to the following processes: the particle did not attempt to move; the particle in site $r$ previously moved in direction $\sigma$, attempted to continue in the same direction but site $r+\sigma$ is occupied; the particle in site $r$ previously moved in direction $-\sigma$, attempted to move backwards but site $r+\sigma$ is occupied; the particle previously moved in direction $\sigma$, was in site $r-\sigma$, continued to move in the same direction, and site $r$ was vacant; and the particle previously moved in direction $-\sigma$, was in site $r-\sigma$, moved in the opposite direction, and site $r$ was vacant. We now define
\begin{eqnarray}
&P(r)=P(r,1)+P(r,-1) ,\nonumber\\
&P(r,r')=P(r,1;r')+P(r,-1;r') ,\nonumber\\
&V(r)=P(r,1)-P(r,-1) ,\nonumber\\
&V(r,r')=P(r,1;r')-P(r,-1;r') ,
\end{eqnarray}
where $P(r)$ and $P(r,r')$ are respectively the probability that site $r$ or sites $r$ and $r'$ are occupied, while $V(r)$ and $V(r,r')$ correspond to the ``velocity'' at site $r$. We then find that (\ref{ev1}) becomes
\begin{eqnarray}
\fl\tau\frac{\partial P(r)}{\partial t}=&-P(r)+\frac{1}{2}\left[P(r-1)+P(r+1)\right]+\delta\left[V(r-1)-V(r+1)\right]+\nonumber\\
&+\frac{1}{2}\left[P(r,r+1)+P(r,r-1)-P(r-1,r)-P(r+1,r)\right]+\nonumber\\
&+\delta\left[V(r,r+1)-V(r,r-1)-V(r-1,r)+V(r+1,r)\right] ,\nonumber\\
\fl\tau\frac{\partial V(r)}{\partial t}=&-V(r)+\frac{1}{2}\left[P(r-1)-P(r+1)\right]+\delta\left[V(r-1)+V(r+1)\right]+\nonumber\\
&+\frac{1}{2}\left[P(r,r+1)-P(r,r-1)-P(r-1,r)+P(r+1,r)\right]+\nonumber\\
&+\delta\left[V(r,r+1)+V(r,r-1)-V(r-1,r)-V(r+1,r)\right] .\label{ev2b}
\end{eqnarray}
Note that $P(r,r')=P(r',r)$ and (\ref{ev2b}) are simplified to
\begin{eqnarray}
\fl\tau\frac{\partial P(r)}{\partial t}=&-P(r)+\frac{1}{2}\left[P(r-1)+P(r+1)\right]+\delta\left[V(r-1)-V(r+1)\right]+\nonumber\\
&+\delta\left[V(r,r+1)-V(r,r-1)-V(r-1,r)+V(r+1,r)\right] ,\nonumber\\
\fl\tau\frac{\partial V(r)}{\partial t}=&-V(r)+\frac{1}{2}\left[P(r-1)-P(r+1)\right]+\delta\left[V(r-1)+V(r+1)\right]+\nonumber\\
&+\left[P(r,r+1)-P(r,r-1)\right]+\nonumber\\
&+\delta\left[V(r,r+1)+V(r,r-1)-V(r-1,r)-V(r+1,r)\right] .\label{ev3}
\end{eqnarray}
We now introduce the three correlation functions $C_{0}(r),C_{\pm}(r)$
\begin{eqnarray}
\fl C_{+}(r)=V(r,r+1)+V(r+1,r)-\left[V(r)P(r+1)+V(r+1)P(r)\right]\nonumber\\
\fl=2\left[P\left(r,+;r+1;+\right)-P\left(r,-;r+1,-\right)\right]-\left[V(r)P(r+1)+V(r+1)P(r)\right] ,\nonumber\\
\fl C_{0}(r)=P(r,r+1)-P(r)P(r+1) ,\nonumber\\
\fl C_{-}(r)=V(r,r+1)-V(r+1,r)-\left[V(r)P(r+1)-V(r+1)P(r)\right]\nonumber\\
\fl=2\left[P\left(r,+;r+1;-\right)-P\left(r,-;r+1,+\right)\right]-\left[V(r)P(r+1)-V(r+1)P(r)\right] ,
\end{eqnarray}
where $P(r,\sigma;r',\sigma')$ is the probability that site $r$ is occupied with a particle that last moved in direction $\sigma$ and site $r'$ is occupied with a particle that last moved in direction $\sigma'$. The evolution equations now read
\begin{eqnarray}
\fl\tau\frac{\partial P(r)}{\partial t}=-P(r)+\frac{1}{2}\left[P(r-1)+P(r+1)\right]+\delta\left[V(r-1)-V(r+1)\right]\left[1-P(r)\right]+\nonumber\\
\fl+\delta V(r)\left[P(r+1)-P(r-1)\right]+\delta\left[C_{+}(r)-C_{+}(r-1)\right], \nonumber\\
\fl\tau\frac{\partial V(r)}{\partial t}=-\left\{1-\delta\left[P(r+1)+P(r-1)\right]\right\}V(r)+\left[\frac{1}{2}-P(r)\right]\left[P(r-1)-P(r+1)\right]+\nonumber\\
\fl+\delta\left[1-P(r)\right]\left[V(r-1)+V(r+1)\right]+C_{0}(r)-C_{0}(r-1)+\delta\left[C_{-}(r)-C_{-}(r-1)\right] .\label{ev4b}
\end{eqnarray}
Note that physically, the correlation $C_{0}$ is just the correlation between the occupancy of the two sites, $C_{+}$ is related to the two particle moving in the same direction, and $C_{-}$ is related to the two particles moving in opposite directions.

We now take the hydrodynamic limit, by assuming that the distance between two adjacent sites, $a$, is very small, and that the mean time between steps, $\tau$, scales as $a^{2}$. Expanding (\ref{ev4b}) to second order in $a$ yields
\begin{eqnarray}
\tau\frac{\partial P}{\partial t}=&\frac{\partial}{\partial x}\left[\frac{a^{2}}{2}\frac{\partial P}{\partial x}-2a\delta V\left(1-P\right)+a\delta C_{+}\right], \nonumber\\
\tau\frac{\partial V}{\partial t}=&-\left(1-2\delta\right)V-a\left[\left(1-2P\right)\frac{\partial P}{\partial x}-\frac{\partial}{\partial x}\left(C_{0}+\delta C_{-}\right)\right]+\nonumber\\
&+a^{2}\left[\frac{\partial^{2}P}{\partial x^{2}}V+\left(1-P\right)\frac{\partial^{2}V}{\partial x^{2}}\right] .
\end{eqnarray}
Since we do not yet consider the extreme case of total persistence in which $\delta=\frac{1}{2}$, and since $P$ is finite, we find that $V$ and $C_{+}$ must scale at most as $a$, and thus
\begin{eqnarray}
V=\frac{a}{1-2\delta}\left[\left(1-2P\right)\frac{\partial P}{\partial x}-\frac{\partial}{\partial x}\left(C_{0}+\delta C_{-}\right)\right] .
\end{eqnarray}
Hence, $P$ satisfies the diffusion equation
\begin{eqnarray}
\frac{\partial P}{\partial t}=\frac{\partial}{\partial x}\left[D\left(\rho\right)\frac{\partial P}{\partial x}+v\left(\rho\right)P\right] ,\label{diff_eq_1s}
\end{eqnarray}
where the diffusion coefficient $D\left(\rho\right)$ and the drift term $v\left(\rho\right)$ are
\begin{eqnarray}
&D\left(\rho\right)=\frac{a^{2}}{2\tau}\left\{1+\frac{4\delta}{1-2\delta}\left(1-P\right)\left[1-2P-\frac{\partial}{\partial\rho}\left(C_{0}+\delta C_{-}\right)\right]\right\} ,\nonumber\\
&v\left(\rho\right)=\frac{2a\delta}{\tau}\frac{C_{+}}{P} .\label{diff_eq_1sb}
\end{eqnarray}

Note that as this is not a gradient model, using the density dependence of the correlation functions at pseudo-equilibrium is only an approximation \cite{Teomy2017}. From symmetry, we find that at pseudo-equilibrium $C_{+}=0$.

\subsection{General memory in $d$ dimensions}
We now consider walkers with a finite general isotropic memory term. We denote by $\eta$ the memory of the particle, and by $\mathbf{\eta}_{n}$ the $n$'th previous step, such that $\mathbf{\eta}_{1}$ is the last step made. The probability that a particle with memory $\eta'$ attempts to move such that its new memory is $\eta$ is given by the matrix element ${\cal M}_{\eta,\eta'}$. The probability that site $\mathbf{r}$ is occupied by a particle with history $\eta$, $P(\mathbf{r},\eta)$ is governed by the evolution equation
\begin{eqnarray}
\fl\tau\frac{\partial P(\mathbf{r},\eta)}{\partial t}=-P(\mathbf{r},\eta)+\sum_{\eta'}P(\mathbf{r},\eta';\mathbf{r}+\mathbf{\eta}_{1}){\cal M}_{\eta,\eta'}+\nonumber\\
\fl+\sum_{\eta'}{\cal M}_{\eta,\eta'}\left[P(\mathbf{r}-\mathbf{\eta}_{1},\eta')-P(\mathbf{r}-\mathbf{\eta}_{1},\eta';\mathbf{r})\right] .\label{genhist_ddim0}
\end{eqnarray}
We now define the state vectors $\textbf{P}(\mathbf{r})$ and $\textbf{P}(\mathbf{r},\mathbf{r}')$ whose components are respectively $P(\mathbf{r},\eta)$ and $P(\mathbf{r},\eta;\mathbf{r}')$, such that Eq. (\ref{genhist_ddim0}) may be written in matrix form as
\begin{eqnarray}
\fl\tau\frac{\partial\textbf{P}(\mathbf{r})}{\partial t}=-\textbf{P}(\mathbf{r})+\sum_{\mathbf{d}}{\cal M}^{\mathbf{d}}\textbf{P}(\mathbf{r},\mathbf{r}+\mathbf{d})+\sum_{\mathbf{d}}{\cal M}^{-\mathbf{d}}\textbf{P}(\mathbf{r},\mathbf{r}-\mathbf{d})+\nonumber\\
\fl+\sum_{\mathbf{d}}{\cal M}^{\mathbf{d}}\left[\textbf{P}(\mathbf{r}-\mathbf{d})-\textbf{P}(\mathbf{r}-\mathbf{d};\mathbf{r})\right]+\sum_{\mathbf{d}}{\cal M}^{-\mathbf{d}}\left[\textbf{P}(\mathbf{r}+\mathbf{d})-\textbf{P}(\mathbf{r}+\mathbf{d};\mathbf{r})\right] ,\label{genhist_ddim1}
\end{eqnarray}
where $\mathbf{d}$ is a unit vector in the $d$ direction, and where ${\cal M}^{\pm\mathbf{d}}$ is the $2^{m}\times 2^{m}$ matrix whose elements are
\begin{eqnarray}
{\cal M}^{\pm\mathbf{d}}_{\eta,\eta'}={\cal M}_{\eta,\eta'}\delta_{\mathbf{\eta}_{1},\pm\mathbf{d}} .
\end{eqnarray}
We now write $\textbf{P}(\mathbf{r})$ and $\textbf{P}(\mathbf{r},\mathbf{r}')$ as a linear combination of the eigenvectors of ${\cal M}$
\begin{eqnarray}
&\textbf{P}(\mathbf{r})=\sum_{n}A_{n}(\mathbf{r})\textbf{V}_{n} ,\nonumber\\
&\textbf{P}(\mathbf{r},\mathbf{r}')=\sum_{n}A_{n}(\mathbf{r},\mathbf{r}')\textbf{V}_{n} ,
\end{eqnarray}
which satisfy
\begin{eqnarray}
{\cal M}\textbf{V}_{n}=\lambda_{n}\textbf{V}_{n} .
\end{eqnarray}
Multiplying (\ref{genhist_ddim1}) from the left by the left eigenvectors of ${\cal M}$, $\textbf{U}^{T}_{n}$, which satisfy $\textbf{U}^{T}_{n}\textbf{V}_{m}=\delta_{m,n}$, yields
\begin{eqnarray}
\fl\tau\frac{\partial A_{n}(\mathbf{r})}{\partial t}=-A_{n}(\mathbf{r})+\sum_{\mathbf{d}}\frac{\lambda_{n}}{2d}\sum_{\sigma=\pm1}\left[A_{n}(\mathbf{r}+\sigma\mathbf{d})+A_{n}(\mathbf{r},\mathbf{r}+\sigma\mathbf{d})-A_{n}(\mathbf{r}+\sigma\mathbf{d},\mathbf{r})\right]-\nonumber\\
\fl-\sum_{\mathbf{d}}\sum_{m\neq n}\sum_{\sigma=\pm1}\sigma\mu_{n,m}\left[A_{m}(\mathbf{r}+\sigma\mathbf{d})-A_{m}(\mathbf{r},\mathbf{r}+\sigma\mathbf{d})-A_{m}(\mathbf{r}+\sigma\mathbf{d};\mathbf{r})\right] ,\label{genhist_ddim2}
\end{eqnarray}
where we defined for brevity
\begin{eqnarray}
\mu_{n,m}=\textbf{U}^{T}_{n}{\cal M}^{\mathbf{d}}\textbf{V}_{m} ,
\end{eqnarray}
which does not depend on $\mathbf{d}$ due to isotropy, and used the relations
\begin{eqnarray}
&\textbf{U}^{T}_{n}{\cal M}^{-\mathbf{d}}\textbf{V}_{m}=\textbf{U}^{T}_{n}\left(\frac{1}{d}{\cal M}-{\cal M}^{\mathbf{d}}\right)\textbf{V}_{m}=\frac{\lambda_{n}}{d}\delta_{m,n}-\mu_{n,m} ,\nonumber\\
&\mu_{n,n}=\frac{\lambda_{n}}{2d} .
\end{eqnarray}
We now introduce the correlation functions
\begin{eqnarray}
\fl C^{\pm}_{n,\mathbf{d}}(\mathbf{r})=A_{n}(\mathbf{r},\mathbf{r}+\mathbf{d})\pm A_{n}(\mathbf{r}+\mathbf{d},\mathbf{r})-\left[A_{n}(\mathbf{r})P(\mathbf{r}+\mathbf{d})\pm A_{n}(\mathbf{r}+\mathbf{d})P(\mathbf{r})\right] ,\label{cordef}
\end{eqnarray}
such that (\ref{genhist_ddim2}) becomes
\begin{eqnarray}
\fl\tau\frac{\partial A_{n}(\mathbf{r})}{\partial t}=-A_{n}(\mathbf{r})+\frac{\lambda_{n}}{2d}\sum_{\mathbf{d}}\sum_{\sigma=\pm1}\left[A_{n}(\mathbf{r}+\sigma\mathbf{d})+A_{n}(\mathbf{r})P(\mathbf{r}+\sigma\mathbf{d})-A_{n}(\mathbf{r}+\sigma\mathbf{d})P(\mathbf{r})\right]\nonumber\\
\fl-\sum_{\mathbf{d}}\sum_{m\neq n}\sum_{\sigma=\pm1}\sigma\mu_{n,m}\left[A_{m}(\mathbf{r}+\sigma\mathbf{d})-A_{m}(\mathbf{r})P(\mathbf{r}+\sigma\mathbf{d})-A_{m}(\mathbf{r}+\sigma\mathbf{d})P(\mathbf{r})\right]\nonumber\\
\fl+\frac{\lambda_{n}}{d}\sum_{\mathbf{d}}\left[C^{-}_{n,\mathbf{d}}(\mathbf{r})-C^{-}_{n,\mathbf{d}}(\mathbf{r}-\mathbf{d})\right]+\sum_{m\neq n}\sum_{\mathbf{d}}\mu_{n,m}\left[C^{+}_{m,\mathbf{d}}(\mathbf{r})-C^{+}_{m,\mathbf{d}}(\mathbf{r}-\mathbf{d})\right] ,\label{genhist_ddim3}
\end{eqnarray}

So far we made no approximations, just transformed the evolution equation into a nicer form. We now take the hydrodynamic limit such that (\ref{genhist_ddim3}) transforms into
\begin{eqnarray}
\fl\tau\frac{\partial A_{n}}{\partial t}=\left(1-\lambda_{n}\right)A_{n}+\lambda_{n}\frac{a^{2}}{2d}\sum_{\mathbf{d}}\left[\left(1-P\right)\frac{\partial^{2}A_{n}}{\partial\mathbf{d}^{2}}+A_{n}\frac{\partial^{2}P}{\partial\mathbf{d}^{2}}\right]\nonumber\\
\fl-2a\sum_{\mathbf{d}}\sum_{m\neq n}\mu_{n,m}\frac{\partial}{\partial{\mathbf{d}}}\left[\left(1-P\right)A_{m}\right]+\lambda_{n}\frac{2a}{d}\sum_{\mathbf{d}}\frac{\partial C^{-}_{n,\mathbf{d}}}{\partial\mathbf{d}}+2a\sum_{m\neq n}\sum_{\mathbf{d}}\mu_{n,m}\frac{\partial C^{+}_{m,\mathbf{d}}}{\partial{\mathbf{d}}} .\label{genhist_hydro}
\end{eqnarray}
Now note that ${\cal M}$ is a reducible stochastic matrix, and thus one of its eigenvalue is unity, $\lambda_{1}=1$, while the real part of the others is strictly smaller than $1$. Furthermore, after a sufficiently long time, the distribution of the states reaches the steady state, which is given by the eigenvector $\textbf{V}_{1}$, and thus $A_{1}=P$. Therefore, we find that for $n>1$, $A_{n}$ scales as $a$, and is thus given by
\begin{eqnarray}
A_{n>1}=\frac{2a\mu_{n,1}}{1-\lambda_{n}}\sum_{\mathbf{d}}\frac{\partial}{\partial\mathbf{d}}\left[(1-P)P-C^{+}_{1,\mathbf{d}}-\frac{\lambda_{n}}{d\mu_{n,1}}C^{-}_{n,\mathbf{d}}\right] ,
\end{eqnarray}
such that $P$ is governed by the drift-diffusion equation
\begin{eqnarray}
\frac{\partial P}{\partial t}=\sum_{\mathbf{d},\mathbf{d}'}\frac{\partial}{\partial\mathbf{d}}\left[D_{\mathbf{d},\mathbf{d}'}\left(P\right)\frac{\partial P}{\partial\mathbf{d}'}+v_{\mathbf{d}}(P)P\right] ,
\end{eqnarray}
with
\begin{eqnarray}
\fl D_{\mathbf{d},\mathbf{d}'}(P)=\frac{a^{2}}{2d\tau}\left[\delta_{\mathbf{d},\mathbf{d}'}-4\sum_{m>1}\frac{\mu_{1,m}\mu_{m,1}}{1-\lambda_{m}}\left(1-P\right)\left(1-2P-\frac{\partial C^{+}_{1,\mathbf{d}'}}{\partial P}-\frac{\lambda_{n}}{d\mu_{m,1}}\frac{\partial C^{-}_{m,\mathbf{d}'}}{\partial P}\right)\right] ,\nonumber\\
\fl v_{\mathbf{d}}(P)=\frac{2a}{\tau P}\sum_{m>1}C^{+}_{m,\mathbf{d}} .
\end{eqnarray}

Regarding the correlations, note that by (\ref{cordef}), $C^{+}_{1,\mathbf{d}}$ is
\begin{eqnarray}
C^{+}_{1,\mathbf{d}}(\mathbf{i})=2\left[P(\mathbf{i},\mathbf{i}+\mathbf{d})-P(\mathbf{i})P(\mathbf{i}+\mathbf{d})\right] ,
\end{eqnarray}
i.e. it is the correlation between the occupancy of different sites. The more complicated term, $C^{-}_{n,\mathbf{d}}$ may also be written as
\begin{eqnarray}
\fl C^{-}_{n,\mathbf{d}}(\mathbf{i})=\nonumber\\
\fl=\sum_{m\neq n}A_{n,m}(\mathbf{i},\mathbf{i}+\mathbf{d})-A_{n,m}(\mathbf{i}+\mathbf{d},\mathbf{i})-\left[A_{n}(\mathbf{i})A_{m}(\mathbf{i}+\mathbf{d})-A_{n}(\mathbf{i}+\mathbf{d})A_{m}(\mathbf{i})\right] ,
\end{eqnarray}
which encodes the asymmetry between two adjacent sites having different histories.

\section{Correlations}
\label{app_corr}
In this section we derive the low-density approximation of the two-point correlations for a general $d$-dimensional model. We start by considering the evolution equation of $P\left(\mathbf{r},\eta;\mathbf{r}',\eta'\right)$, which is the probability that site $\mathbf{r}$ is occupied by a particle with memory $\eta$ and site $\mathbf{r}'\neq\mathbf{r}$ is occupied by a particle with memory $\eta'$. This probability evolves according to
\begin{eqnarray}
\fl\tau\frac{\partial P\left(\mathbf{r},\eta;\mathbf{r}',\eta'\right)}{\partial t}=-2P\left(\mathbf{r},\eta;\mathbf{r'},\eta'\right)\nonumber\\
\fl+\sum_{\eta''}{\cal M}_{\eta,\eta''}\left[\delta_{\mathbf{r}',\mathbf{r}+\eta_{1}}P\left(\mathbf{r},\eta'';\mathbf{r}',\eta'\right)+\left(1-\delta_{\mathbf{r}',\mathbf{r}+\eta_{1}}\right)P\left(\mathbf{r},\eta'';\mathbf{r}',\eta';\mathbf{r}+\eta_{1}\right)\right]\nonumber\\
\fl+\sum_{\eta''}{\cal M}_{\eta',\eta''}\left[\delta_{\mathbf{r},\mathbf{r}'+\eta'_{1}}P\left(\mathbf{r},\eta;\mathbf{r}',\eta''\right)+\left(1-\delta_{\mathbf{r},\mathbf{r}'+\eta'_{1}}\right)P\left(\mathbf{r},\eta;\mathbf{r}',\eta'';\mathbf{r}'+\eta'_{1}\right)\right]\nonumber\\
\fl+\sum_{\eta''}{\cal M}_{\eta,\eta''}\left(1-\delta_{\mathbf{r}',\mathbf{r}-\eta_{1}}\right)\left[P\left(\mathbf{r}-\eta_{1},\eta'';\mathbf{r}',\eta'\right)-P\left(\mathbf{r}-\eta_{1},\eta'';\mathbf{r}',\eta';\mathbf{r}\right)\right]\nonumber\\
\fl+\sum_{\eta''}{\cal M}_{\eta',\eta''}\left(1-\delta_{\mathbf{r},\mathbf{r}'-\eta'_{1}}\right)\left[P\left(\mathbf{r},\eta;\mathbf{r}'-\eta'_{1},\eta''\right)-P\left(\mathbf{r},\eta;\mathbf{r}'-\eta'_{1},\eta'';\mathbf{r}'\right)\right] ,\label{eq_corr1}
\end{eqnarray}
where $\eta_{1}$ is the last step in the memory $\eta$, and $P\left(\mathbf{r},\eta;\mathbf{r}',\eta';\mathbf{r}''\right)$ is the probability that site $\mathbf{r}$ is occupied by a particle with memory $\eta$, site $\mathbf{r}'$ is occupied by a particle with memory $\eta'$, and site $\mathbf{r}''$ is occupied by a particle with any memory. We assume implicitly that all three sites are different. The first term in (\ref{eq_corr1}) accounts for the case when both particles do not move, the second (third) term describes the attempt by the particle in site $\mathbf{r}$ ($\mathbf{r}'$) to move to an already occupied site, and the fourth (fifth) term describes a successful move to site $\mathbf{r}$ ($\mathbf{r}'$).

In order to have a close set of equations, we consider the following approximation for the three-point correlations
\begin{eqnarray}
&P\left(\mathbf{r},\eta;\mathbf{r}',\eta;\mathbf{r}''\right)\approx\left(1-\delta_{\mathbf{r},\mathbf{r}'}\right)\left(1-\delta_{\mathbf{r},\mathbf{r}''}\right)\left(1-\delta_{\mathbf{r}',\mathbf{r}''}\right)\nonumber\\
&\frac{1}{3}\left[P\left(\mathbf{r},\eta;\mathbf{r}',\eta'\right)P\left(\mathbf{r}''\right)+P\left(\mathbf{r},\eta;\mathbf{r}''\right)P\left(\mathbf{r}',\eta'\right)+P\left(\mathbf{r}',\eta';\mathbf{r}''\right)P\left(\mathbf{r},\eta\right)\right] ,
\end{eqnarray}
where the extra Kronecker delta functions are needed to keep the approximation equal to zero if two of the sites are the same.
In the steady state the one-point functions are known
\begin{eqnarray}
&P\left(\mathbf{r}\right)=\rho ,\nonumber\\
&P\left(\mathbf{r},\eta\right)=\rho P_{ss}\left(\eta\right) ,
\end{eqnarray}
where $P_{ss}\left(\eta\right)$ is the steady state probability of a particle to be with memory $\eta$. Furthermore, due to translational invariance, the two-point functions depend only on the distance between the two sites. We therefore define
\begin{eqnarray}
P_{2}\left(\mathbf{r},\eta,\eta'\right)\equiv P\left(\mathbf{r}+\mathbf{\Delta r},\eta;\mathbf{\Delta r},\eta'\right) 
\end{eqnarray}
for any $\mathbf{\Delta r}$. The subscript $2$ reminds us that this is a two-point function. In the steady state we set the temporal derivative to zero, and find that Eq. (\ref{eq_corr1}) may be approximated by
\begin{eqnarray}
\fl0=-2P_{2}\left(\mathbf{r},\eta,\eta'\right)+\frac{\rho}{3}\left[{\cal M}_{\eta,\eta''}P_{2}\left(\mathbf{r},\eta'',\eta'\right)+{\cal M}_{\eta',\eta''}P_{2}\left(\mathbf{r},\eta,\eta''\right)\right]-\nonumber\\
\fl-\frac{\rho}{3}\sum_{\eta'',\eta'''}P_{ss}\left(\eta''\right)\left[{\cal M}_{\eta,\eta''}P_{2}\left(\mathbf{r},\eta''',\eta'\right)+{\cal M}_{\eta',\eta''}P_{2}\left(\mathbf{r},\eta,\eta'''\right)\right]+\nonumber\\
\fl+\left(1-\frac{\rho}{3}\right)\sum_{\eta''}\left[{\cal M}_{\eta,\eta''}P_{2}\left(\mathbf{r}-\eta_{1},\eta'',\eta'\right)+{\cal M}_{\eta',\eta''}P_{2}\left(\mathbf{r}+\eta'_{1},\eta,\eta''\right)\right]+\nonumber\\
\fl+\frac{\rho}{3}\sum_{\eta'',\eta'''}P_{ss}\left(\eta''\right)\left[{\cal M}_{\eta,\eta''}P_{2}\left(\mathbf{r}+\eta_{1},\eta''',\eta'\right)+{\cal M}_{\eta',\eta''}P_{2}\left(\mathbf{r}-\eta'_{1},\eta,\eta'''\right)\right]+\nonumber\\
\fl+\left(1-\frac{\rho}{3}\right)\sum_{\eta''}\left[\delta_{\mathbf{r},-\eta_{1}}{\cal M}_{\eta,\eta''}P_{2}\left(-\eta_{1},\eta'',\eta'\right)+\delta_{\mathbf{r},\eta'_{1}}{\cal M}_{\eta',\eta''}P_{2}\left(\eta'_{1},\eta,\eta''\right)\right]+\nonumber\\
\fl+\frac{\rho}{3}\sum_{\eta'',\eta'''}P_{ss}\left(\eta''\right)\left[\delta_{\mathbf{r},\eta_{1}}{\cal M}_{\eta,\eta''}P_{2}\left(-\eta_{1},\eta',\eta'''\right)+\delta_{\mathbf{r},-\eta'_{1}}{\cal M}_{\eta',\eta''}P_{2}\left(-\eta'_{1},\eta,\eta'''\right)\right]+\nonumber\\
\fl+\frac{\rho}{3}P_{ss}\left(\eta'\right)\left(\delta_{\mathbf{r},\eta_{1}}-\delta_{\mathbf{r},-\eta_{1}}\right)\sum_{\eta'',\eta'''}{\cal M}_{\eta,\eta''}P_{2}\left(-\eta_{1},\eta'',\eta'''\right)+\nonumber\\
\fl+\frac{\rho}{3}P_{ss}\left(\eta\right)\left(\delta_{\mathbf{r},-\eta'_{1}}-\delta_{\mathbf{r},\eta'_{1}}\right)\sum_{\eta'',\eta'''}{\cal M}_{\eta',\eta''}P_{2}\left(-\eta'_{1},\eta'',\eta'''\right) ,
\end{eqnarray}
where we used
\begin{eqnarray}
P_{2}\left(\mathbf{r},\eta,\eta'\right)=P_{2}\left(-\mathbf{r},\eta',\eta\right) .
\end{eqnarray}
Also, by the definition of the steady state we have
\begin{eqnarray}
\sum_{\eta'}{\cal M}_{\eta,\eta'}P_{ss}(\eta')=P_{ss}(\eta) ,
\end{eqnarray}
and thus
\begin{eqnarray}
\fl0=-2P_{2}\left(\mathbf{r},\eta,\eta'\right)+\frac{\rho}{3}\left[{\cal M}_{\eta,\eta''}P_{2}\left(\mathbf{r},\eta'',\eta'\right)+{\cal M}_{\eta',\eta''}P_{2}\left(\mathbf{r},\eta,\eta''\right)\right]-\nonumber\\
\fl-\frac{\rho}{3}\sum_{\eta''}\left[P_{ss}\left(\eta\right)P_{2}\left(\mathbf{r},\eta'',\eta'\right)+P_{ss}\left(\eta'\right)P_{2}\left(\mathbf{r},\eta,\eta''\right)\right]+\nonumber\\
\fl+\left(1-\frac{\rho}{3}\right)\sum_{\eta''}\left[{\cal M}_{\eta,\eta''}P_{2}\left(\mathbf{r}-\eta_{1},\eta'',\eta'\right)+{\cal M}_{\eta',\eta''}P_{2}\left(\mathbf{r}+\eta'_{1},\eta,\eta''\right)\right]+\nonumber\\
\fl+\frac{\rho}{3}\sum_{\eta''}\left[P_{ss}\left(\eta\right)P_{2}\left(\mathbf{r}+\eta_{1},\eta'',\eta'\right)+P_{ss}\left(\eta'\right)P_{2}\left(\mathbf{r}-\eta'_{1},\eta,\eta''\right)\right]+\nonumber\\
\fl+\left(1-\frac{\rho}{3}\right)\sum_{\eta''}\left[\delta_{\mathbf{r},-\eta_{1}}{\cal M}_{\eta,\eta''}P_{2}\left(-\eta_{1},\eta'',\eta'\right)+\delta_{\mathbf{r},\eta'_{1}}{\cal M}_{\eta',\eta''}P_{2}\left(\eta'_{1},\eta,\eta''\right)\right]+\nonumber\\
\fl+\frac{\rho}{3}\sum_{\eta''}\left[P_{ss}\left(\eta\right)\delta_{\mathbf{r},\eta_{1}}P_{2}\left(\eta_{1},\eta'',\eta'\right)+P_{ss}\left(\eta'\right)\delta_{\mathbf{r},-\eta'_{1}}P_{2}\left(-\eta'_{1},\eta,\eta''\right)\right]+\nonumber\\
\fl+\frac{\rho}{3}P_{ss}\left(\eta'\right)\left(\delta_{\mathbf{r},\eta_{1}}-\delta_{\mathbf{r},-\eta_{1}}\right)\sum_{\eta'',\eta'''}{\cal M}_{\eta,\eta''}P_{2}\left(-\eta_{1},\eta'',\eta'''\right)+\nonumber\\
\fl+\frac{\rho}{3}P_{ss}\left(\eta\right)\left(\delta_{\mathbf{r},-\eta'_{1}}-\delta_{\mathbf{r},\eta'_{1}}\right)\sum_{\eta'',\eta'''}{\cal M}_{\eta',\eta''}P_{2}\left(\eta'_{1},\eta''',\eta''\right) .
\end{eqnarray}
In matrix form this may be written as
\begin{eqnarray}
0={\cal Q}_{1}\textbf{P}_{2}\left(\mathbf{r}\right)+\sum_{\sigma\mathbf{d}}{\cal Q}^{\sigma\mathbf{d}}_{2}\textbf{P}_{2}\left(\mathbf{r}-\sigma\mathbf{d}\right)+\sum_{\sigma\mathbf{d}}{\cal Q}^{\sigma\mathbf{d}}_{3}\delta_{\mathbf{r},-\sigma\mathbf{d}}\textbf{P}_{2}\left(\mathbf{r}\right) ,
\end{eqnarray}
with
\begin{eqnarray}
&{\cal Q}_{1}=-2{\cal I}+\frac{\rho}{3}\left({\cal M}_{1}+{\cal M}_{2}-{\cal S}_{1}-{\cal S}_{2}\right) ,\nonumber\\
&{\cal Q}^{\sigma\mathbf{d}}_{2}=\left(1-\frac{\rho}{3}\right)\left({\cal M}^{\sigma\mathbf{d}}_{1}+{\cal M}^{-\sigma\mathbf{d}}_{2}\right)+\frac{\rho}{3}\left({\cal S}^{-\sigma\mathbf{d}}_{1}+{\cal S}^{\sigma\mathbf{d}}_{2}\right) ,\nonumber\\
&{\cal Q}^{\sigma\mathbf{d}}_{3}={\cal Q}^{\sigma\mathbf{d}}_{2}-\frac{\rho}{3}\left[{\cal S}_{2}\left({\cal M}^{\sigma\mathbf{d}}_{1}-{\cal M}^{-\sigma\mathbf{d}}_{1}\right)+{\cal S}_{1}\left({\cal M}^{-\sigma\mathbf{d}}_{2}-{\cal M}^{\sigma\mathbf{d}}_{2}\right)\right] ,\nonumber\\
\end{eqnarray}
The subscript $1$ and $2$ on the matrices ${\cal M}$ and ${\cal S}$ denotes whether they act on particle $1$ or $2$, the matrices ${\cal S}^{\sigma\mathbf{d}}$ are defined by
\begin{eqnarray}
\left[{\cal S}^{\pm\mathbf{d}}\right]_{\eta,\eta'}=P_{ss}\left(\eta\right)\delta_{\eta_{1},\pm\mathbf{d}} ,
\end{eqnarray}
and the matrix ${\cal S}$ is
\begin{eqnarray}
{\cal S}=\sum_{\sigma\mathbf{d}}{\cal S}^{\sigma\mathbf{d}} .
\end{eqnarray}

In order to solve the recursion equation we define the generating function $\textbf{G}\left(\mathbf{\theta}\right)$
\begin{eqnarray}
\textbf{G}\left(\mathbf{\theta}\right)=\sum_{\mathbf{r}}e^{i\mathbf{r}\cdot\mathbf{\theta}}\textbf{P}_{2}\left(\mathbf{r}\right) .
\end{eqnarray}
Multiplying the recursion equation by $e^{i\mathbf{r}\cdot\mathbf{\theta}}$ and summing over $\mathbf{r}$ yields an equation on $\textbf{G}\left(\mathbf{\theta}\right)$
\begin{eqnarray}
0={\cal Q}_{1}\textbf{G}\left(\mathbf{\theta}\right)+\sum_{\sigma\mathbf{d}}{\cal Q}^{\sigma\mathbf{d}}_{2}e^{i\sigma\mathbf{d}\cdot\mathbf{\theta}}\textbf{G}\left(\mathbf{\theta}\right)+\nonumber\\
+\sum_{\sigma\mathbf{d}}{\cal Q}^{\sigma\mathbf{d}}_{3}e^{-i\sigma\mathbf{d}\cdot\mathbf{\theta}}\textbf{P}_{2}\left(-\sigma\mathbf{d}\right)-\sum_{\sigma\mathbf{d}}{\cal Q}^{\sigma\mathbf{d}}_{2}\textbf{P}_{2}\left(-\sigma\mathbf{d}\right) ,
\end{eqnarray}
where the last term comes from summing over also the non-existent equation for $\mathbf{r}=\mathbf{0}$ which needs to be removed. Rearranging the equation we get
\begin{eqnarray}
\fl\left({\cal Q}_{1}+\sum_{\sigma\mathbf{d}}{\cal Q}^{\sigma\mathbf{d}}_{2}e^{i\sigma\mathbf{d}\cdot\mathbf{\theta}}\right)\textbf{G}\left(\mathbf{\theta}\right)=\sum_{\sigma\mathbf{d}}\left({\cal Q}^{\sigma\mathbf{d}}_{2}\textbf{P}_{2}-{\cal Q}^{\sigma\mathbf{d}}_{3}e^{-i\sigma\mathbf{d}\cdot\mathbf{\theta}}\right)\textbf{P}_{2}\left(-\sigma\mathbf{d}\right) .
\end{eqnarray}
Inverting the Fourier transform yields
\begin{eqnarray}
\fl\textbf{P}_{2}\left(\mathbf{r}\right)=\frac{1}{\left(2\pi\right)^{d}}\int_{\mathbf{\theta}\in\left[0,2\pi\right]^{d}}e^{-i\mathbf{r}\cdot\mathbf{\theta}}\textbf{G}\left(\mathbf{\theta}\right)=\nonumber\\
\fl\frac{1}{\left(2\pi\right)^{d}}\int_{\mathbf{\theta}\in\left[0,2\pi\right]^{d}}e^{-i\mathbf{r}\cdot\mathbf{\theta}}\left({\cal Q}_{1}+\sum_{\sigma\mathbf{d}}{\cal Q}^{\sigma\mathbf{d}}_{2}e^{i\sigma\mathbf{d}\cdot\mathbf{\theta}}\right)^{-1}\sum_{\sigma\mathbf{d}}\left({\cal Q}^{\sigma\mathbf{d}}_{2}-{\cal Q}^{\sigma\mathbf{d}}_{3}e^{-i\sigma\mathbf{d}\cdot\mathbf{\theta}}\right)\textbf{P}_{2}\left(-\sigma\mathbf{d}\right) .
\end{eqnarray}
Setting $\mathbf{r}=\sigma'\mathbf{d}'$ yields
\begin{eqnarray}
\textbf{P}_{2}\left(\sigma'\mathbf{d}'\right)=\frac{1}{\left(2\pi\right)^{d}}\int_{\mathbf{\theta}\in\left[0,2\pi\right]^{d}}e^{-i\sigma'\mathbf{d}'\cdot\mathbf{\theta}}\left({\cal Q}_{1}+\sum_{\sigma\mathbf{d}}{\cal Q}^{\sigma\mathbf{d}}_{2}e^{i\sigma\mathbf{d}\cdot\mathbf{\theta}}\right)^{-1}\times\nonumber\\
\times\sum_{\sigma\mathbf{d}}\left({\cal Q}^{\sigma\mathbf{d}}_{2}-{\cal Q}^{\sigma\mathbf{d}}_{3}e^{-i\sigma\mathbf{d}\cdot\mathbf{\theta}}\right)\textbf{P}_{2}\left(-\sigma\mathbf{d}\right) .
\end{eqnarray}
This is a set of linear equations between the $2d$ vectors $\textbf{P}_{2}\left(\sigma\mathbf{d}\right)$, which may be written as $\textbf{P}_{2}={\cal N}\textbf{P}_{2}$. Hence, $\textbf{P}_{2}\left(\sigma\mathbf{d}\right)$ is found by finding the unit eigenvalue of the matrix ${\cal N}$. Since the eigenvector is found up to a multiplicative constant, we use another boundary condition that at $\left|\mathbf{r}\right|\rightarrow\infty$ the two sites are uncorrelated and thus the elements of $\textbf{P}_{2}(\infty)$ are given by $P_{ss}(\eta)P_{ss}(\eta')$ where $P_{ss}(\eta)$ is the probability that a particle with history $\eta$ is in the steady state.

For one-dimensional systems, there is another, simpler way to derive the correlations. Consider the recursion equation reduced to one dimension
\begin{eqnarray}
0={\cal Q}_{1}\textbf{P}_{2}(r)+\sum_{\sigma}{\cal Q}^{\sigma}_{2}\textbf{P}_{2}(r-\sigma)+\sum_{\sigma}{\cal Q}^{\sigma}_{3}\delta_{r,-\sigma}\textbf{P}_{2}(r) .
\end{eqnarray}
Since $\textbf{P}_{2}(0)=0$ by definition, the recursion equation for $r>0$ is independent of the $\textbf{P}_{2}(r)$ for $r<0$. Hence, without loss of generality we consider only $r>0$. The recursion equation reads
\begin{eqnarray}
\fl\textbf{P}_{2}(r+1)=-\left[\left({\cal Q}^{-}_{2}\right)^{-1}{\cal Q}_{1}+\delta_{r,1}\left({\cal Q}^{-}_{2}\right)^{-1}{\cal Q}^{-}_{3}\right]\textbf{P}_{2}(r)-\left({\cal Q}^{-}_{2}\right)^{-1}{\cal Q}^{+}_{2}\textbf{P}_{2}(r-1) ,
\end{eqnarray}
This may also be written as
\begin{eqnarray}
\fl\left(\begin{array}{c}\textbf{P}_{2}(r+1)\\\textbf{P}_{2}(r)\end{array}\right)=\nonumber\\
\fl=-\left(\begin{array}{cc}\left({\cal Q}^{-}_{2}\right)^{-1}{\cal Q}_{1}+\delta_{r,1}\left({\cal Q}^{-}_{2}\right)^{-1}{\cal Q}^{-}_{3}&\left({\cal Q}^{-}_{2}\right)^{-1}{\cal Q}^{+}_{2}\\-{\cal I}&0\end{array}\right)\left(\begin{array}{c}\textbf{P}_{2}(r)\\\textbf{P}_{2}(r-1)\end{array}\right) .
\end{eqnarray}
The solution is
\begin{eqnarray}
\fl\left(\begin{array}{c}\textbf{P}_{2}(r+1)\\\textbf{P}_{2}(r)\end{array}\right)=\left(-1\right)^{r}\left(\begin{array}{cc}\left({\cal Q}^{-}_{2}\right)^{-1}{\cal Q}_{1}&\left({\cal Q}^{-}_{2}\right)^{-1}{\cal Q}^{+}_{2}\\-{\cal I}&0\end{array}\right)^{r-1}\nonumber\\
\fl\left(\begin{array}{cc}\left({\cal Q}^{-}_{2}\right)^{-1}{\cal Q}_{1}+\left({\cal Q}^{-}_{2}\right)^{-1}{\cal Q}^{-}_{3}&\left({\cal Q}^{-}_{2}\right)^{-1}{\cal Q}^{+}_{2}\\-{\cal I}&0\end{array}\right)\left(\begin{array}{c}\textbf{P}_{2}(1)\\0\end{array}\right) .
\end{eqnarray}
The normalisation is taken from the requirement that at $r\rightarrow\infty$ the correlations decay to zero and the vector $\textbf{P}_{2}(\infty)$ is equal to the steady state distribution of two uncorrelated sites.

Taking for example the simplest case, a one-dimensional system with one-step memory, we find after straightforward but cumbersome calculations that
\begin{eqnarray}
\textbf{P}_{2}(r)=\frac{\rho^{2}}{4}\left(\begin{array}{c}1\\1\\1\\1\end{array}\right)+x^{r}_{1}\textbf{X}_{1}+x^{r}_{2}\textbf{X}_{2} ,
\end{eqnarray}
where $\textbf{X}_{1}$ and $\textbf{X}_{2}$ are vectors whose exact dependence on $\rho$ and $\delta$ is too cumbersome to write explicitly, and $x_{1}$ and $x_{2}$ are
\begin{eqnarray}
\fl x_{1}=\frac{6-2\delta\rho-\sqrt{3\left(1-2\delta\right)\left(9+6\delta-4\delta\rho\right)}}{3+6\delta-2\delta\rho} ,\nonumber\\
\fl x_{2}=\frac{9+18\delta-24\delta\rho+4\delta\rho^{2}-\sqrt{3\left(3+6\delta-4\delta\rho\right)\left(9+18\delta-36\delta\rho+8\delta\rho^{2}\right)}}{4\delta\rho\left(3-\rho\right)} .
\end{eqnarray}
For small densities, we expand $\textbf{P}_{2}(r)$ to second order in $\rho$ and find that the two-point correlations are
\begin{eqnarray}
\fl C_{+,+}(r)=C_{-,-}(r)=\frac{1}{2}\left[C_{+,-}(r)+C_{-,+}(r)\right]=\frac{\rho^{2}}{4}\frac{1-8\delta-4\delta^{2}+\left(1+2\delta\right)^{2}x_{0}}{\left(1-2\delta\right)\left(1-4\delta-4\delta^{2}\right)}x^{r}_{0} ,\nonumber\\
\fl\frac{1}{2}\left[C_{+,-}(r)-C_{-,+}(r)\right]=\frac{\rho^{2}}{2}\frac{-1+x_{0}}{1-4\delta-4\delta^{2}}x^{r}_{0} ,
\end{eqnarray}
with
\begin{eqnarray}
x_{0}=\frac{2-\sqrt{4-\left(1+2\delta\right)^{2}}}{1+2\delta} .
\end{eqnarray}

\end{document}